\documentclass[a4paper,11pt]{article}
\usepackage{jheppub} % for details on the use of the package, please see the JINST-author-manual
\usepackage{lineno}
\usepackage{amsmath}
\usepackage{physics}
\usepackage{subfigure}
\usepackage{mathrsfs}
\usepackage{ulem}
\def\bc{\begin{center}}

\def\ec{\end{center}}
\def\be{\begin{eqnarray}}
\def\ee{\end{eqnarray}}

\def\blue{\color{blue}}

\newcommand{\Ket}[1]{\left |#1\right )}
\newcommand{\Bra}[1]{\left (#1\right |}
\newcommand{\overlap}[2]{(#1|#2)}
\title{\boldmath Krylov Complexity in the Schr\"odinger Field Theory}

\author[a]{Peng-Zhang He}
\author[a,b]{, Hai-Qing Zhang}
\affiliation[a]{Center for Gravitational Physics, Department of Space Science, Beihang University, Beijing 100191, China}
\affiliation[b]{Peng Huanwu Collaborative Center for Research and Education, Beihang University, Beijing 100191, China}

\emailAdd{hepzh@buaa.edu.cn}
\emailAdd{hqzhang@buaa.edu.cn}

\abstract{{
		We investigate the Krylov complexity of Schr\"odinger field theories, focusing on both bosonic and fermionic systems within the grand canonical ensemble that includes a chemical potential. Krylov complexity measures operator growth in quantum systems by analyzing how operators spread within the Krylov space, a subspace of the Hilbert space spanned by successive applications of the superoperator $[H,\cdot]$ on an initial operator. Using the Lanczos algorithm, we construct an orthonormal Krylov basis and derive the Lanczos coefficients, which govern the operator connectivity and thus characterize the complexity. Our study reveals that the Lanczos coefficients $\{b_{n}\}$ are independent of the chemical potential, while $\{a_{n}\}$ exhibits a dependence on it. Both $\{a_{n}\}$ and $\{b_{n}\}$ show linear relationships with respect to $n$. For both bosonic and fermionic systems, the Krylov complexities behave similarly over time, especially at late times, due to the analogous profiles of the squared absolute values of their autocorrelation functions $\abs{\varphi_{0}(t)}^{2}$. The Krylov complexity grows exponentially with time, but its asymptotic scaling factor $\lambda_{K}$ is significantly smaller than the twice of the slope of the $\{b_{n}\}$ coefficients, contrasting to the relativistic field theories where the scaling aligns more closely with the twice of the slope of $\{b_{n}\}$. }}

\begin{document}
\maketitle
\flushbottom
%\linenumbers
   \section{Introduction}
   The complexity of quantum systems refers to the increasingly intricate nature of their states and operations over time. Recent studies on quantum complexity in Krylov space, known as Krylov complexity, have attracted significant attention. This area of research spans multiple fields, including quantum many-body physics \cite{Parker:2018yvk,Rabinovici:2022beu,Liu:2022god,Trigueros:2021rwj,Bhattacharjee:2022ave,Caputa:2022yju,Bhattacharjee:2023uwx,Craps:2024suj}, quantum field theory \cite{Camargo:2022rnt,He:2024xjp,Chattopadhyay:2024pdj,Malvimat:2024vhr,Vasli:2023syq,Kundu:2023hbk,Avdoshkin:2022xuw,Khetrapal:2022dzy,Adhikari:2022whf,Dymarsky:2021bjq}, holographic theories \cite{Rabinovici:2023yex}, random matrix theory \cite{Kar:2021nbm}, and cosmology \cite{Adhikari:2022oxr,Li:2024ljz,Li:2024iji,Li:2024kfm}. For a comprehensive review, please refer to \cite{Nandy:2024htc} and the associated references. Since the pioneering work published in \cite{Parker:2018yvk}, Krylov complexity has become essential for revealing quantum dynamics' information-theoretic aspects.
   
    Krylov complexity was developed to analyze quantum systems' integrable and chaotic behaviors as they approach the thermodynamic limit. Investigations suggest potential correlations between Krylov complexity and physical quantities, including the out-of-time-order correlator (OTOC) \cite{Hashimoto:2017oit}, the eigenstate thermalization hypothesis (ETH) \cite{PhysRevE.50.888}, entanglement entropy \cite{PhysRevE.99.032213}, and Nielsen complexity \cite{PhysRevD.100.046020}. People find Krylov complexity to be a well-defined quantity with a clear physical interpretation\cite{Parker:2018yvk,He:2024xjp,Lv:2023jbv}, and it is easier to compute than other complexities. Specifically, Krylov complexity describes how a wave function spreads through a Hilbert space, corresponding to operators' growth in the Heisenberg picture, within a special basis known as the Krylov basis. In simpler terms, this wave function's (average) position is the Krylov complexity, which we will elaborate on in Section \ref{sec2.1}. Furthermore, Krylov complexity is utilized in the Schr\"odinger picture, referred to as spread complexity \cite{Balasubramanian:2022tpr}, to describe the evolution of quantum states. Krylov complexity has become an active field of research with discoveries emerging continuously. For more information, see \cite{Ganguli:2024uiq,Fu:2024fdm,Nandy:2024mml,Fan:2024iop,Xu:2024gfm,Caputa:2024sux,Baggioli:2024wbz,Rabinovici:2020ryf,Bhattacharjee:2022vlt,Rabinovici:2021qqt,Caputa:2022eye,Bhattacharya:2023zqt,Erdmenger:2023wjg,Bhattacharjee:2022qjw,Bhattacharya:2022gbz,Bhattacharjee:2022lzy,Camargo:2023eev,Huh:2023jxt,Camargo:2024deu,He:2022ryk,Caputa:2024vrn,Hornedal:2022pkc,Bento:2023bjn,Nandy:2023brt}.

    In Ref. \cite{Parker:2018yvk}, the authors hypothesize that the Lanczos coefficients $\{b_n\}$ should grow as rapidly as possible in chaotic quantum systems. The results indicate that Lanczos coefficients' maximum possible growth rate is linear (with logarithmic corrections in one dimension). Generally, OTOC grows exponentially with large time as $e^{\lambda_L t}$ ($\lambda_L$ is the Lyapunov exponent), indicating the rapid information scrambling and the sensitivity to the initial conditions. In contrast, Krylov complexity also increases exponentially with large time {as $e^{\lambda_K t}$}, where $\lambda_K$ is dubbed the Krylov exponent \cite{Parker:2018yvk}. It was conjectured that twice the linear growth rate of the Lanczos coefficients (equivalent to the Krylov exponent) could be an upper limit of the quantum Lyapunov exponent $\lambda_{L}$, i.e., $\lambda_{L}\le \lambda_{K}$.  Although the operator growth hypothesis effectively distinguishes chaotic and non-chaotic systems in lattice models, it was later found ineffective for quantum systems with infinite degrees of freedom, such as quantum field theories (QFTs). It turns out that even in integrable quantum field theories like the free scalar field, the Krylov complexity still exhibits exponential growth \cite{Dymarsky:2019elm,Dymarsky:2021bjq}. This implies that many subtle aspects of Krylov complexity may not be fully understood in the context of quantum field theory.
   
   In this paper, we will investigate Krylov complexity within the framework of Schr\"odinger field theory \cite{altland2010condensed}.  The development of Schr\"odinger field theory began in 1926 when Austrian physicist Erwin Schr\"odinger proposed the Schr\"odinger equation, which became the foundation of quantum mechanics. Schr\"odinger field theory, as a non-relativistic field theory, was subsequently used to describe quantum fields that follow the Schr\"odinger equation, especially in the context of many-body systems and situations where the number of particles changes. Over time, Schr\"odinger field theory has been widely applied in Bose-Einstein condensates, the Bogolyubov–de Gennes equations for superconductors, superfluids, and many-body theory, and has become an indispensable part of modern physics \cite{harris2014pedestrian,Mintchev:2022xqh}. The Schr\"odinger field operator is not a Hermitian operator; hence, it is more troublesome to handle compared to the Krylov complexity of the free scalar field \cite{Camargo:2022rnt,He:2024xjp}. This is because there is no relationship $O^\dagger = O$; at this time, the Lanczos coefficients have two sequences $\{a_n\}$ and $\{b_n\}$, and when dealing with two-point functions, the simplification of calculations using the Hermitian operator cannot be utilized. This paper studies the Krylov complexity in bosonic and fermionic cases with the grand canonical ensemble, as the chemical potential is less than or equal to zero. We find that the Lanczos coefficients $\{a_n\}$ and $\{b_n\}$ are both linear with respect to $n$. Interestingly, $\{b_n\}$ is independent of the chemical potentials while $\{a_n\}$ depends on it. Regardless of the bosonic or the fermionic case, the behavior of Krylov complexities is always similar, which is due to the similar behaviors of the square of the absolute values for the auto-correlation function $\lvert\varphi_0(t)\rvert^2$.  In the late times, the Krylov complexity has approached exponential growth. Interestingly, we find that the asymptotic exponential scaling of the Krylov complexity is always smaller than twice the slope of $\{b_n\}$, which differs from the relativistic free scalar field where the asymptotic exponential scaling of the Krylov complexity is roughly twice the slope of $\{b_n\}$. We attribute this discrepancy to the existence of another Lanczos coefficient $\{a_n\}$ in the non-relativistic Schr\"odinger field theory.

%   This paper is organized as follows. In Section \ref{sec2}, we will present the theoretical foundations of Krylov complexity, including the construction of the Krylov basis, the computation of Lanczos coefficients, and the relationship between the Wightman function and the spectral function. In Section \ref{sec3}, we will provide an introduction to Schr\"odinger field theory, discussing how to derive the spectral function and present the Wightman power spectrum. In Section \ref{sec4}, we will calculate the Lanczos coefficients for both bosonic and fermionic cases in the Schr\"odinger field theory and analyze the behavior of Krylov complexity under various chemical potentials. Section \ref{sec5} will provide a summary of this paper.
   
   { Our work systematically investigates the features of Krylov complexity in the non-relativistic Schr\"odinger field theories. The logical progression unfolds as follows: \begin{itemize}
   		\item Section \ref{sec2} establishes the operator growth framework in Krylov space, detailing the construction of orthonormal bases through Gram-Schmidt orthogonalization and deriving the recursive relations for Lanczos coefficients $\{a_{n}\}$ and $\{b_{n}\}$. We further introduce the complexity dynamics in quantum field theory by connecting the Wightman power spectrum to the spectral densities and then introduce the moment method as a computational tool for the Lanczos coefficients.  
   		\item Based on the foundations in Section \ref{sec2}, Section \ref{sec3} specializes in the Schr\"odinger field theory. We analytically solve the spectral function $\rho(\omega,\mathbf{k}) = 2\pi\delta(\xi_{\mathbf{k}}-\omega)$ through the thermal propagators, which explicitly demonstrates how the chemical potential $\mu$ constrains Wightman power spectrum via the Heaviside $\Theta(\mu-\omega)$ cutoff. The derived Wightman power spectrum $f^W(\omega) = \text{thermal factor}\times(\mu-\omega)^{(d-3)/2}$ is normalized in five  dimensions $(d=5)$ of the spacetime, in which the thermal factor has distinct forms for bosonic and fermionic operators.  
   		\item Section \ref{sec4} unveils the universal growth patterns of the Lanczos coefficients and the Krylov complexity: Lanczos coefficients exhibit the linear scaling $\beta a_{n}\approx -4(n+1) - \mu$ and $\beta b_{n} \approx 2n + 1$, demonstrating the $\mu$-independence of $\{b_{n}\}$ while the $\mu$-dependence of $\{a_{n}\}$. Numerical solutions to the discrete Schr\"odinger equation confirm the exponential growth of Krylov complexity $K(t) \sim e^{\lambda_K t}$, with $\lambda_K \approx 2.746/\beta$ significantly below the $2\alpha$ limit in the relativistic free scalar field, highlighting the suppression role of $\{a_{n}\}$ terms. A conceptual analogy to the Lie-algebraic models is provided to understand these findings. 
   		\item Section \ref{sec5} provides a summary of this paper. 
   		\item Technical foundations are consolidated in the appendices: Appendix \ref{appa} details the Runge-Kutta implementation for the evolution of $\varphi_{n}(t)$, Appendix \ref{appB} rigorously derives the $\delta$-function spectral structure. In Appendix \ref{appc}, we discuss the conditions under which the Lanczos coefficients do not exhibit staggering.
   \end{itemize} }
   
   \section{Preliminaries}\label{sec2}
   To ensure the reader's understanding and make the paper self-contained, we will provide a detailed introduction to the basics of Krylov complexity in this section. { This section outlines the theoretical framework for computing Krylov complexity. We first introduce the construction of Krylov basis and Lanczos coefficients through the Gram-Schmidt orthogonalization process. Next, we use the moment method to establish the relationship between the Wightman power spectrum, moments, and the Lanczos coefficients. Finally, we connect the spectral function to the Wightman two-point function, which is essential for our field theory computations in the subsequent sections.}
   \subsection{Krylov complexity}\label{sec2.1}
   In the quantum theory, a Heisenberg operator $O(t)$ satisfies the Heisenberg equation \cite{sakurai2020modern}
   \begin{equation}\label{2.1}
   	\partial_{t}O(t)=i[H,O(t)],
   \end{equation}
   where $H$ is the time-independent Hamiltonian. The operator $O(t)$ may be either Hermitian or non-Hermitian. We can define a superoperator $\mathcal{L} \equiv [H, \cdot]$ such that 
   \begin{equation}
   	\mathcal{L}O(t) \equiv [H, O(t)].
   \end{equation}
   Then we have 
   \begin{equation}
   	\partial_{t}O(t)=i\mathcal{L}O(t).
   \end{equation}
   This expression resembles the Schr\"odinger equation for a state, which has a formal solution:
   \begin{equation}
   	O(t)=e^{i\mathcal{L}t}\mathcal{O}(0)=\sum_{n=0}^{\infty}\frac{(it)^{n}}{n!}\mathcal{L}^{n}O,
   \end{equation}
   where $O\equiv O(0)$ and
   \begin{equation}
\mathcal{L}^{0}O=O,\qquad   	\mathcal{L}O=[H,O],\qquad \mathcal{L}^{2}O=[H,[H,O]],\qquad\cdots.
   \end{equation}
   The operator $O(t)$ belongs to a subspace of operators known as the Krylov space, which is spanned by $\{\mathcal{L}^{n}O\},n=0,1,2,\cdots$. The Krylov space is also a Hilbert space, and one can define an inner product for it. One can define the Hilbert space vector as $\Ket{A}$ corresponding to operator $A$, then the inner product can be formulated as \cite{Parker:2018yvk,Caputa:2021sib} 
   \begin{equation}\label{16}
   	\overlap{A}{B}^{g}_{\beta}=\int_{0}^{\beta}g(\lambda)\expval{e^{\lambda H}A^{\dagger}e^{-\lambda H}B}_{\beta}d\lambda.
   \end{equation}
   In this formula, $\beta$ represents the inverse temperature, while $\expval{\cdots}_{\beta}$ is the thermal expectation value
   \begin{equation}
   	\expval{A}_{\beta}=\frac{1}{Z}\tr(e^{-\beta H}A), \quad Z=\tr(e^{-\beta H}).
   \end{equation} 
   For \eqref{16} to be a suitable inner product, the function $g(\lambda)$ should satisfy
   \begin{equation}
   	g(\lambda)\ge0,\qquad g(\beta-\lambda)=g(\lambda),\qquad \frac{1}{\beta}\int_{0}^{\beta}d\lambda g(\lambda)=1.
   \end{equation}
   It can be observed that 
   \begin{equation}
   	\expval{e^{\lambda H}A^{\dagger}e^{-\lambda H}\mathcal{L}B}_{\beta}=\expval{e^{-\lambda H}(\mathcal{L}A)^{\dagger}e^{-\lambda H}B}_{\beta},
   \end{equation}
   indicating that $\mathcal{L}$ is a Hermitian operator in Krylov space. Assuming that $O$ is initially normalized, then
   \begin{equation}
   	\expval{e^{\lambda H}O^{\dagger}(t)e^{-\lambda H}O(t)}_{\beta}=\expval{e^{\lambda H}O^{\dagger}(0)e^{-\lambda H}O(0)}_{\beta}.
   \end{equation}
    This means that regardless of whether the operator $O$ is Hermitian, an initial normalized operator will always remain normalized as it evolves. 
   
   In the Krylov space, $\{\mathcal{L}^{n}O\}$ is the basis, as we mentioned above, but such a basis is typically neither orthogonal nor normalized. We aim to construct an orthonormal basis, known as the Krylov basis $\{\Ket{O_{n}}\}$, which can be constructed from the Gram-Schmidt orthogonalization process. The Gram-Schmidt orthogonalization process is a method that involves the following procedures \cite{geroch2013quantum}. Let $\xi,\eta_{1},\eta_{2},\cdots,\eta_{n}$ be elements of the Hilbert space $\mathcal{H}$. Then
   \begin{equation}\label{2.11}
   	 \xi=u_{1}\eta_{1}+\cdots +u_{n}\eta_{n}+\tau,\qquad (u_{i}\in\mathbb{C}, {\rm and}~(\eta_{i},\tau)=0),
   \end{equation}
   where $(\eta,\tau)$ denotes the inner product of $\eta$ and $\tau$ in the Hilbert space, and $\tau$ is also a vector in Hilbert space. That is to say, any vector in $\mathcal{H}$ can be written as a linear combination of $\eta_{1},\cdots ,\eta_{n}$, plus a vector $\tau$ perpendicular to the $\eta$'s. In the Krylov space, it is clear that there is 
   \begin{equation}\label{12}
   	\mathcal{L}\Ket{O_{n}}=\sum_{m=0}^{n}{\alpha^{n}}_{m}\Ket{O_{m}}+b_{n+1}\Ket{O_{n+1}},
   \end{equation}
   where ${\alpha^n}_{m}=\Bra{O_{m}}\mathcal{L}\Ket{O_{n}}$, $b_{n+1}=\Bra{O_{n+1}}\mathcal{L}\Ket{O_{n}}$. If $\{{\alpha^{n}}_{m}\}$ and $\{b_{n}\}$ are real numbers, then
   \begin{equation}
   	b_{n+1}={\alpha^{n}}_{n+1}={\alpha^{n+1}}_{n}.
   \end{equation} 
   It can be observed that
   \begin{equation}
   	\Bra{O_{m}}\mathcal{L}\Ket{O_{n}}=   	\Bra{O_{n}}\mathcal{L}\Ket{O_{m}}={\alpha^{n}}_{m}={\alpha^{m}}_{n}=0,\qquad \text{for}\  m-n\ge2 ~\text{and}~ n\ge0.
   \end{equation}
    Specifically, Let $\Ket{O_{0}}=\Ket{O}$ be a { normalized} operator, then
   \begin{equation}
   	\mathcal{L}\Ket{O_{0}}=a_{0}\Ket{O_{0}}+b_{1}\Ket{O_{1}}.
   \end{equation}
   In accordance with the notation used in \eqref{12}, we define $a_{0}={\alpha^{0}}_{0}=\Bra{O_{0}}\mathcal{L}\Ket{O_{0}}$. Subsequently, we can set
   \begin{equation}
   	a_{n}={\alpha^{n}}_{n}=\Bra{O_{n}}\mathcal{L}\Ket{O_{n}}. 
   \end{equation}
Following the previous operations, we have
\begin{eqnarray}
	\mathcal{L}\Ket{O_{1}}&=&b_{1}\Ket{O_{0}}+a_{1}\Ket{O_{1}}+b_{2}\Ket{O_{2}},\\
	\mathcal{L}\Ket{O_{2}}&=&b_{2}\Ket{O_{1}}+a_{2}\Ket{O_{2}}+b_{3}\Ket{O_{3}},\\
	&\cdots&\nonumber\\
	\mathcal{L}\Ket{O_{n}}&=&b_{n}\Ket{O_{n-1}}+a_{n}\Ket{O_{n}}+b_{n+1}\Ket{O_{n+1}},\\
	&\cdots&\nonumber
\end{eqnarray}
These two sequences $\{b_{n}\}$ and $\{a_{n}\}$ are called the Lanczos coefficients. In the Krylov space, the operator $O(t)$ can be expressed as \cite{Parker:2018yvk}
\begin{equation}\label{2.20}
	\Ket{O(t)}=\sum_{n=0}^{\infty}i^{n}\varphi_{n}(t)\Ket{O_{n}},
\end{equation}
where $\varphi_{n}(t)=\overlap{O_{n}}{O(t)}$. It is not difficult to see that $\varphi_{n}(0)=\delta_{n0}$ and $\varphi_{n}(t)=0$ for $n<0$. The vector $\left|O(t)\right)$ is always normalized, which is equivalent to
\begin{equation}\label{2.21}
	\sum_{n}\abs{\varphi_{n}(t)}^{2}=1.
\end{equation}
By substituting equation \eqref{2.20} into the Heisenberg equation \eqref{2.1}, we can express the left side of equation \eqref{2.1} as follows:
\begin{equation}
	\partial_{t}\Ket{O(t)}=\sum_{n=0}^{\infty}i^{n}\partial_{t}\varphi_{n}(t)\Ket{O_{n}},
\end{equation} 
and the right side becomes:
\begin{equation}
	\begin{aligned}
		i\mathcal{L}\Ket{O(t)}&=\sum_{n=0}^{\infty}i^{n+1}\varphi_{n}(t)\mathcal{L}\Ket{O_{n}}\\
		&=\sum_{n=0}^{\infty}i^{n+1}\varphi_{n}(t)\left[b_{n}\Ket{O_{n-1}}+a_{n}\Ket{O_{n}}+b_{n+1}\Ket{O_{n+1}}\right]\\
		&=\sum_{n=0}^{\infty}i^{n}[i^{2}\varphi_{n+1}b_{n+1}+ia_{n}\varphi_{n}(t)+b_{n}\varphi_{n-1}(t)]\Ket{O_{n}},
	\end{aligned}
\end{equation}
where we have assumed that $b_{0}=0$. By matching the coefficients of $\Ket{O_{n}}$, we derive the discrete Schr\"odinger equation
\begin{equation}\label{2.23}
	\partial_{t}\varphi_{n}(t)=ia_{n}\varphi_{n}(t)+b_{n}\varphi_{n-1}(t)-b_{n+1}\varphi_{n+1}(t).
\end{equation}
The evolution of operators over time, or their growth, can now be understood as a hopping problem on a one-dimensional chain. In this paper, we define the Krylov complexity as follows \cite{Dymarsky:2021bjq}:
\begin{equation}\label{2.24}
	K(t)=1+\sum_{n=0}^{\infty}n\abs{\varphi_{n}(t)}^{2},
\end{equation}
in order to be consistent with previous literature in quantum field theory \cite{He:2024xjp}. It is important to emphasize that operators's growth is directly related to an increase in the number of contributing $n$. However, this does not imply that increased operators will consistently increase the Krylov complexity. Detailed discussions on this topic can be found in \cite{He:2024xjp}.

To obtain the Krylov complexity, one must first derive the Lanczos coefficients and solve the discrete Schr\"odinger equation \eqref{2.23}. The Lanczos coefficients can be calculated using the moment method \cite{viswanath1994recursion}, where the moments are defined as follows:
\begin{equation}\label{2.25}
	\mu_{n}\equiv\Bra{O}\mathcal{L}^{n}\Ket{O}=\Bra{O}\left .(-i)^{n}\frac{d^{n}}{dt^{n}}\Ket{O(t)}\right |_{t=0}=(-i)^{n}\left .\frac{d^{n}}{dt^{n}}\varphi_{0}(t)\right |_{t=0}.
\end{equation}
In the second equality, we used the Heisenberg equation. To compute the moments, one can introduce an important quantity known as the auto-correlation function
\begin{equation}\label{2.26}
	\begin{aligned}
		C(t):&=\overlap{O(t)}{O}=\varphi_{0}^{\ast}(t),
	\end{aligned}
\end{equation}
and the Wightman two-point function, 
\begin{equation}
\begin{aligned}
		\Pi^{W}(t)&=\varphi_{0}(t)
		=\overlap{O}{O(t)}
		=\expval{e^{\beta H/2}O^{\dagger}(0)e^{-\beta H/2}O(t)}_{\beta}\\
		&=\expval{O^{\dagger}(0)e^{iH(t+i\beta/2)}O(0)e^{-iH(t+i\beta /2)}}_{\beta}\\
		&=\expval{O^{\dagger}(0)O(t+i\beta/2)}_{\beta}.
\end{aligned}
\end{equation}
Here, we have defined the inner product using the Wightman inner product
\begin{equation}\label{2.28}
	\overlap{A}{B}=\expval{e^{H\beta/2}A^{\dagger}e^{-H\beta/2}B}_{\beta},
\end{equation}
which means taking $ g(\lambda)=\delta(\lambda-\beta/2)$ in Eq.\eqref{16}.  The autocorrelation function \eqref{2.26} can also be written as:
	\begin{equation}
		C(t)=\expval{O^\dagger(t-i\beta/2)O(0)}_{\beta}.
	\end{equation}
The Fourier transform of the auto-correlation function is referred to as the Wightman power spectrum
\begin{equation}\label{2.29}
	f^{W}(\omega)=\int_{-\infty}^{\infty}dte^{i\omega t}C(t).
\end{equation}
The moments \eqref{2.25} then can be rewritten in terms of $f^{W}(\omega)$ as,
\begin{equation}\label{2.30}
	\mu_{n}=\frac{1}{2\pi}\int_{-\infty}^{\infty}d\omega \omega^{n}[f^{W}(\omega)]^{\ast}.
\end{equation}
Since \( \mu_0 = 1 \), \( f^{W}(\omega) \) should satisfy the normalization condition
\begin{equation}\label{2.31}
	1=\frac{1}{2\pi}\int_{-\infty}^{\infty}d\omega [f^{W}(\omega)]^{\ast}.
\end{equation}
Once we have obtained the moments, we can calculate the Lanczos coefficients through the following recurrence relations \cite{viswanath1994recursion}
\begin{eqnarray}
	M^{(n)}_{k}&=&L^{(n-1)}_{k}-L^{(n-1)}_{n-1}\frac{M^{(n-1)}_{k}}{M^{(n-1)}_{n-1}},\\
	L^{(n)}_{k}&=&\frac{M^{(n)}_{k+1}}{M^{(n)}_{n}}-\frac{M^{(n-1)}_{k}}{M^{(n-1)}_{n-1}},
\end{eqnarray}
where $n=1,\cdots,2K$, $k=n,\cdots,2K-n+1$ and 
\begin{equation}
	M^{(0)}_{k}=(-1)^{k}\mu_{k},\qquad L^{(0)}_{k}=(-1)^{k+1}\mu_{k+1},\qquad k=0,\cdots,2K.
\end{equation}
The resulting Lanczos coefficients are 
\begin{equation}
	b_{n}^{2}=M^{(n)}_{n},\qquad a_{n}=-L^{(n)}_{n},\qquad n=0,\cdots,K.
\end{equation}
These equations may seem complicated, but they are not difficult to operate in practice. We just need to regard $M^{(n)}_{k}$ and $L^{(n)}_{k}$ as two matrices, $k$ as the row index, and $n$ as the column index. Then, once we have calculated the $n$-th column of $M$, we can calculate the $n$-th column of $L$. After obtaining these coefficients, we can use the Runge-Kutta method in the time direction to solve the discrete Schr\"odinger equation to obtain $\varphi_n(t)$ (see Appendix \ref{appa}). Consequently, we can compute $K(t)$ based on the definition of Krylov complexity.
{ \subsubsection{Linear Lanczos coefficients in chaos systems and free field theories}

In Ref.\cite{Parker:2018yvk}\footnote{ There are vanishing $a_{n}$ coefficients in \cite{Parker:2018yvk}.}, it is conjectured that for chaotic quantum many-body systems, the Lanczos coefficients will exhibit asymptotic linearity, namely
\begin{equation}\label{2.38}
	b_{n}\sim \alpha n+\gamma,\qquad n\rightarrow \infty.
\end{equation}
This is referred to as the universal operator growth hypothesis. Many evidences have shown that this conjecture is valid in various models \cite{Parker:2018yvk,Nandy:2024htc}. The asymptotic slope $\alpha $ of $b_{n}$ is an important quantity, since the exponential decay rate of the Wightman power spectrum $f^W(\omega)\sim e^{-\abs{\omega}/\omega_{0}}$ is closely related to $\alpha$, such that $\omega_{0}=(2/\pi)\alpha$. Besides, $\pm i\pi/(2\alpha)$ are the poles of the auto-correlation function closest to the origin. In addition, $2\alpha$ is the exponential growth rate of the Krylov complexity, and $2\alpha$ is an upper bound of the Lyapunov exponent. 

In the field theory, assuming that $b_{n}$ is sufficiently smooth, one will obtain the exponential growth rate of the Krylov complexity as $\lambda_{K}=2\pi/\beta$ \cite{Parker:2018yvk,Avdoshkin:2019trj}, and then this conjecture is reduced to the Maldacena-Shenker-Stanford (MSS) bound \cite{Maldacena:2015waa}
\begin{equation}
	\lambda_{L}\le 2\pi/\beta.
\end{equation}
The universal operator growth hypothesis \eqref{2.38} still holds, and thus, the exponential behavior of Krylov complexity is not exactly a signature of chaos. In Ref.\cite{Dymarsky:2021bjq}, the authors investigated the operator growth of local operators in the Krylov space for various CFT models. They found that in the cases they considered, the asymptotic behavior of $b_{n}$ for large $n$ is $ \beta b_{n}\approx \pi(n+\Delta +1/2)$\footnote{ $\Delta$ is the conformal dimension of the operator.} and the Krylov complexity provides a bound for the OTOC that reduces to the MSS bound. The Krylov complexity exhibits exponential growth in all examples, including free and rational CFTs. This implies that systems with exponential growth of Krylov complexity are not necessarily chaotic. An interesting phenomenon in CFT is that the Lanczos coefficients are divided into two smooth families, one for odd $n$ and one for even $n$. However, the asymptotic behavior of Krylov complexity remains the exponential growth. In the free scalar field theory, Lanczos coefficients with staggering behavior and the Krylov complexity with asymptotic exponential growth are examined in \cite{Camargo:2022rnt,He:2024xjp}. In Ref.\cite{Camargo:2022rnt}, it is pointed out that if the following two conditions are satisfied, the Lanczos coefficients will not exhibit staggering:
\begin{enumerate}
	\item The power spectrum is finite and positive at $\omega=0$, i.e., $0<f^{W}(0)<\infty$;
	\item The derivative of the power spectrum $f^{W}(\omega)$ is a continuous function of $\omega$ for $-\Lambda<\omega <\Lambda$, where $\Lambda $ is a UV cutoff.
\end{enumerate}
In Appendix \ref{appc}, we discuss the conditions under which staggering occurs when the Lanczos coefficients include both $b_{n}$ and $a_{n}$. These discussions consider the effect of the symmetry of the Wightman power spectrum.

}

\subsection{Wightman function and spectral function}
In quantum field theory, spectral functions can represent various correlation functions. We will demonstrate that the auto-correlation function $C(t)$ or Wightman power spectrum $f^{W}(\omega)$ can also be derived from the spectral function. Similar operations can also be referred to in \cite{Camargo:2022rnt}.

Consider two bosonic or fermionic operators, $\phi_{a}$ and $\psi_{b}$, with subscripts used to denote their internal degrees of freedom. We declare that the bosonic operator $\psi_b$ is defined as $\phi_b^\dagger$, while the fermionic operator $\psi_b$ is expressed as $\bar{\phi}_b = \phi^\dagger_b \gamma^0$, where $\gamma^0$ represents the Dirac matrix \cite{peskin2018introduction,kapusta2007finite}. Define the Wightman function as
\begin{gather}
	D^{>}_{ab}(t,\mathbf{x};t^{\prime},\mathbf{x}^{\prime})\equiv\expval{\phi_{a}(t,\mathbf{x})\psi_{b}(t^{\prime},\mathbf{x}^{\prime})}_{\beta},\\
	D^{<}_{ab}(t,\mathbf{x};t^{\prime},\mathbf{x}^{\prime})\equiv\eta\expval{\psi_{b}(t^{\prime},\mathbf{x^{\prime}})\phi_{a}(t,\mathbf{x})}_{\beta}.
\end{gather}
For the bosonic field, $\eta=1$, while for the fermionic field, $\eta=-1$. Using $O(t)=e^{iHt}Oe^{-iHt}$, we have
	\begin{equation}
	\phi_{a}(t+i\beta,\mathbf{x})=e^{-\beta H}\phi_{a}(t,\mathbf{x})e^{\beta H}.
\end{equation}
From this equation, one can demonstrate the Kubo-Martin-Schwinger (KMS) relation in the configuration space as
\begin{equation}
	D^{<}_{ab}(t+i\beta,\mathbf{x};t^{\prime},\mathbf{x}^{\prime})=\eta D_{ab}^{>}(t,\mathbf{x};t^{\prime},\mathbf{x}^{\prime}).
\end{equation}
Assuming that the system has translational symmetries in both time and space directions, we can derive
\begin{gather}
	D^{>}_{ab}(t,\mathbf{x};t^{\prime},\mathbf{x}^{\prime})=D^{>}_{ab}(t-t^{\prime},\mathbf{x}-\mathbf{x}^{\prime}),\\
	D^{<}_{ab}(t,\mathbf{x};t^{\prime},\mathbf{x}^{\prime})=D^{<}_{ab}(t-t^{\prime},\mathbf{x}-\mathbf{x}^{\prime}).
\end{gather}
Without loss of generality, we can set $(t^{\prime} = 0, \mathbf{x}^{\prime} = \mathbf{0})$ and perform a Fourier transform on the Wightman function to obtain
\begin{gather}
	D^{>}_{ab}(\omega,\mathbf{k})=\int_{-\infty}^{\infty}dt\int d^{d-1}\mathbf{x}e^{i(\omega t-\mathbf{k}\cdot\mathbf{x})}D^{>}_{ab}(t,\mathbf{x};0,\mathbf{0}),\\
	D^{<}_{ab}(\omega,\mathbf{k})=\int_{-\infty}^{\infty}dt\int d^{d-1}\mathbf{x}e^{i(\omega t-\mathbf{k}\cdot \mathbf{x})}D^{<}_{ab}(t,\mathbf{x};0,{\mathbf{0}}).
\end{gather}
Applying the KMS relation in configuration space enables the derivation of the KMS relation in momentum space:
\begin{equation}\label{2.43}
	D^{>}_{ab}(\omega,\mathbf{k})=\eta e^{\beta\omega}D^{<}_{ab}(\omega,\mathbf{k}).
\end{equation}
The one can introduce a new correlation function
\begin{equation}\label{2.45}
	\rho_{ab}(t,\mathbf{x};t^{\prime},\mathbf{x}^{\prime})=\expval{[\phi_{a}(t,\mathbf{x}),\psi_{b}(t^{\prime},\mathbf{x}^{\prime})]_{\mp}}_{\beta},
\end{equation}
in which $[\cdot, \cdot]_{\mp}$ takes the commutator for the bosonic field or the anticommutator for the fermionic field. Its Fourier transform is referred to as the spectral density or spectral function
\begin{equation}\label{2.48}
	\rho_{ab}(\omega,\mathbf{k})=D^{>}_{ab}(\omega,\mathbf{k})-D^{<}_{ab}(\omega,\mathbf{k}).
\end{equation}
From \eqref{2.43} and \eqref{2.48}, one can get
\begin{gather}
	D^{>}_{ab}(\omega,\mathbf{k})=\frac{\eta e^{\beta \omega}}{\eta e^{\beta \omega}-1}\rho_{ab}(\omega,\mathbf{k}),\\
	D^{<}_{ab}(\omega,\mathbf{k})=\frac{1}{\eta e^{\beta\omega}-1}\rho_{ab}(\omega,\mathbf{k}).
\end{gather}
To calculate the Krylov complexity of bosonic or fermionic operators, we can define
\begin{equation}\label{2.49}
	\Pi^{W}_{ab}(t,\mathbf{x})=\expval{\psi_{b}(0,\mathbf{0})\phi_{a}(t+i\beta/2,\mathbf{x})}_{\beta}=\eta D_{ab}^{<}(t+i\beta/2,\mathbf{x};0,\mathbf{0}).
\end{equation}
Then we have 
\begin{equation}
	\Pi^{W}(t)=\Pi^{W}_{ab}(t,\mathbf{0}).
\end{equation}
Performing a Fourier transform on \eqref{2.49} yields
\begin{equation}
	\begin{aligned}
		\Pi^{W}_{ab}(\omega,\mathbf{k})&=\eta\int dt\int d^{d-1}\mathbf{x}e^{i(\omega t-\mathbf{k}\cdot\mathbf{x})}D^{<}_{ab}(t+i\beta/2,\mathbf{x};0,\mathbf{0})\\
		&=\eta\int dt\int d^{d-1}\mathbf{x} e^{\beta\omega/2}e^{i(\omega t-\mathbf{k}\cdot\mathbf{x})}D^{<}_{ab}(t,\mathbf{x};0,\mathbf{0})\\
		&=\eta e^{\beta \omega/2}D^{<}_{ab}(\omega,\mathbf{k})\\
	&=\frac{\eta    e^{\frac{\beta  \omega }{2}}}{\eta  e^{\beta  \omega }-1}\rho_{ab}(\omega,\mathbf{k}).
	\end{aligned}
\end{equation}
Then, we can derive the following expression: 
\begin{equation}\label{2.52}
	\begin{aligned}
		f^{W}(\omega)&=\int_{-\infty}^{\infty}dte^{i\omega t}[\Pi^{W}(t)]^{\ast}
		=\int_{-\infty}^{\infty}dt\int d^{d-1}\mathbf{x}e^{i\omega t}[\Pi^{W}_{ab}(t,\mathbf{x})]^{\ast}\delta^{(d-1)}(\mathbf{x}-\mathbf{0})\\
		&=\int_{-\infty}^{\infty}dt\int d^{d-1}\mathbf{x}\int\frac{d\omega^{\prime}}{2\pi}\int\frac{d^{d-1}\mathbf{k}}{(2\pi)^{d-1}}e^{i\omega t}[e^{-i(\omega^{\prime}t-\mathbf{k}\cdot\mathbf{x})}\Pi^{W}_{ab}(\omega^{\prime},\mathbf{k})]^{\ast}\delta^{(d-1)}(\mathbf{x}-\mathbf{0})\\
		&=\int d\omega^{\prime}\int\frac{d^{d-1}\mathbf{k}}{(2\pi)^{d-1}}[\Pi^{W}_{ab}(\omega^{\prime},\mathbf{k})]^{\ast}\delta(\omega+\omega^{\prime})
		=\int\frac{d^{d-1}\mathbf{k}}{(2\pi)^{d-1}}[\Pi^{W}_{ab}(-\omega,\mathbf{k})]^{\ast}\\
		&=\frac{\eta  e^{-\frac{\beta  \omega }{2}}}{\eta  e^{-\beta  \omega }-1}\int\frac{d^{d-1}\mathbf{k}}{(2\pi)^{d-1}}[\rho_{ab}(-\omega,\mathbf{k})]^{\ast}.
	\end{aligned}
\end{equation}
\section{Wightman power spectrum of Schr\"odinger field theory}\label{sec3}
 In the Schr\"odinger field theory, the Lagrangian for non-relativistic free bosons and fermions with mass $m$ is given by
\begin{equation}
	\mathscr{L}=\psi^{\dagger}\left (i\frac{\partial}{\partial t}+\frac{\nabla^{2}}{2m}\right )\psi.
\end{equation}
Verifying that the Euler-Lagrange equations derived from this Lagrangian are indeed the Schr\"odinger equation is straightforward. Suppose $\psi$ is an operator with canonical commutation relations. In that case, it describes a collection of identical non-relativistic bosons, and while $\psi$ is an operator with canonical anti-commutation relations, the field describes identical fermions \cite{Mintchev:2022xqh}. This Lagrangian is symmetric under the following global $U(1)$ transformation
\begin{equation}
	\psi\rightarrow e^{i\alpha}\psi,\qquad \psi^{\dagger}\rightarrow\psi^{\dagger}e^{-i\alpha}.
\end{equation}
The corresponding conserved charge is
\begin{equation}
	N=\int d^{d-1}\mathbf{x}\mathscr{N}=\int d^{d-1}\mathbf{x}\psi^{\dagger}\psi,
\end{equation}
where $\mathscr{N}$ is the charge density. The conjugate field of $\psi$ is
\begin{equation}
	\Pi=\frac{\partial \mathscr{L}}{\partial\dot{\psi}}=i\psi^{\dagger}.
\end{equation}
Then we can write down the Hamiltonian density as
\begin{equation}
	\mathscr{H}=\Pi\dot{\psi}-\mathscr{L}=\psi^{\dagger}\left (-\frac{\nabla^{2}}{2m}\right )\psi.
\end{equation}
In the grand canonical ensemble, the partition function is
\begin{equation}\label{3.6}
	\mathcal{Z}=\tr(e^{-\beta(H-\mu N)}).
\end{equation}
in which $H$ is the Hamiltonian with $H=\int d^{d-1}\mathbf{x}\mathscr{H}$ and $\mu$ is the chemical potential\footnote{ Please do not confuse with the chemical potential $\mu$ with the moments $\mu_i$ in the Eq.\eqref{2.25}}. Consequently, the partition function can be rewritten in the path integral form as
\begin{equation}
	\begin{aligned}
		\mathcal{Z}&=\int\mathcal{D}\Pi\mathcal{D}\psi\exp(\int_{0}^{\beta}d\tau \int d^{d-1}\mathbf{x}(\Pi i\partial_{\tau}\psi-\mathscr{H}+\mu\mathscr{N}))\\
		&=\int\mathcal{D}\psi^{\dagger}\mathcal{D}\psi\exp{\int_{0}^{\beta}d\tau\int d^{d-1}\mathbf{x}(\mathscr{L}+\mu\mathscr{N})},
	\end{aligned}
\end{equation}
where $\tau=it$. The inverse of the thermal propagator of the Schr\"odinger field is
\begin{equation}\label{3.8}
	\mathcal{D}^{-1}(K)\equiv\mathcal{D}^{-1}(\omega_{n},\mathbf{k})=i\omega_{n}+\frac{\mathbf{k}^{2}}{2m}-\mu,
\end{equation}
where $K=(k^{0},\mathbf{k})=(-i\omega_{n},\mathbf{k})$. From the thermal propagator, we can obtain the spectral function as \footnote{Since the situation we are considering does not contain internal degrees of freedom, we can safely omit the subscript ``$ab$'' in the spectral function $\rho_{ab}$ in Eq.\eqref{2.45}. The derivation of Eq.\eqref{3.9} can be found in the Appendix \ref{appB}.}
\begin{equation}\label{3.9}
	\rho(\omega,\mathbf{k})=2\pi\delta(\xi_{\mathbf{k}}-\omega),\qquad \xi_{\mathbf{k}}=\frac{\mathbf{k}^{2}}{2m}-\mu.
\end{equation}
It should be noted that we cannot directly substitute this spectral function into the equation \eqref{2.52} to calculate the Wightman power spectrum because the field $\psi$ here, in the sense of \eqref{2.28}, is not necessarily normalized. Therefore, we can redefine
\begin{equation}
	\rho(\omega,\mathbf{k})=\mathcal{N}\delta(\xi_{\mathbf{k}}-\omega)
\end{equation}
as the spectral function, where $\mathcal{N}$ is a normalization constant that needs to be determined from the equation \eqref{2.31}.

To calculate $f^W(\omega)$, we first compute the integral 
\begin{equation}
	I\equiv\int\frac{d^{d-1}\mathbf{k}}{(2\pi)^{d-1}}[\rho(-\omega,\mathbf{k})]^{\ast}=\mathcal{N}\int\frac{d^{d-1}\mathbf{k}}{(2\pi)^{d-1}}\delta(\xi_{\mathbf{k}}+\omega).
\end{equation}
Using the identity
\begin{equation}
	\int\frac{d^{d}\mathbf{l}}{(2\pi)^{d}}=\int\frac{d\Omega_{d}}{(2\pi)^{d}}\int_{0}^{\infty}dl l^{d-1},\qquad
	\Omega_{d}\equiv\int d\Omega_{d}=\frac{2\pi^{d/2}}{\Gamma(d/2)},
\end{equation}
we have 
\begin{equation}
\begin{aligned}
	I&=	\mathcal{N} \Omega_{d-1}\int_{0}^{\infty}dk\frac{k^{d-2}}{(2\pi)^{d-1}}\delta\left (\frac{k^{2}}{2m}-\mu+\omega\right )\\
	&=	\mathcal{N} \Omega_{d-1}\int_{0}^{\infty}dk\frac{k^{d-2}}{(2\pi)^{d-1}}\sqrt{\frac{m}{2(\mu-\omega)}}\delta(k-\sqrt{2m(\mu-\omega)})\Theta(\mu-\omega)\\
	&=\mathcal{N}\Omega_{d-1}\frac{m^{\frac{d-1}{2}}[2(\mu-\omega)]^{\frac{d-3}{2}}}{(2\pi)^{d-1}}\Theta(\mu-\omega).
\end{aligned}
\end{equation}
in which $\Theta(x)$ is the Heaviside step function with $\Theta(x<0)=0$ and $\Theta(x\geq0)=1$. Then, absorbing all constants into $\mathcal{N}$, we get
\begin{equation}\label{3.14}
	f^{W}(\omega)=\mathcal{N}(\mu,\beta,d,\eta)(\mu-\omega)^{\frac{d-3}{2}}\frac{\eta  e^{-\frac{\beta  \omega }{2}}}{\eta  e^{-\beta  \omega }-1}\Theta(\mu-\omega).
\end{equation}
{ Here, $\mathcal{N}(\mu,\beta,d,\eta)$ is the normalization factor determined by the normalization condition of the Wightman power spectrum \eqref{2.31}.} For simplicity, in this paper, we will only consider the case of five dimensions\footnote{ Our selection of $5$ dimensions $(d=5)$ stems from the fundamental constraints in handling the Wightman power spectrum integral of the form \eqref{3.14}. Specifically:
	\begin{itemize}
		\item Even-dimensional spacetime limitations:\\
		When spacetime dimension is even $(d=2k, k\in Z)$, the term $(\mu-\omega)^{(d-3)/2}$ in the Wightman spectrum produces integrals of the type $(\mu-\omega)^{k-3/2}/\sinh\left (\beta \omega/2\right )$. But in this case, the moments will be hard to be evaluated. The same difficulty is also encountered in \cite{Camargo:2022rnt}.
		\item Odd-dimensional spacetime advantage:\\
		When the spacetime is of odd dimension, the Wightman power spectrum is integrable. But for general spacetime with $ d=2k+1$, we still cannot obtain an analytical expression. Therefore, we choose {\blue a simple} case, $d=5$.
\end{itemize}}. 
Thus, for bosonic fields and fermionic fields, the Wightman power spectrum can be written in the specific form as
\begin{equation}\label{3.15}
	f^{W}(\omega)=\left \{\begin{aligned}
		&\mathcal{N}(\mu,\beta,5,1)\frac{(\omega-\mu)}{2}\frac{1}{\sinh\frac{\beta\omega}{2}}\Theta(\mu-\omega),\qquad \text{bosonic field}\\
		&\mathcal{N}(\mu,\beta,5,-1)\frac{(\mu-\omega)}{2}\frac{1}{\cosh\frac{\beta\omega}{2}}\Theta(\mu-\omega).\qquad \text{fermionic field}
	\end{aligned}\right .
\end{equation}

\section{Lanczos coefficients and Krylov complexity of Schr\"odinger field theory}\label{sec4}
Since the Wightman power spectrum has distinct expressions for bosonic and fermionic fields, as shown in Eq.\eqref{3.15}, we will examine the bosonic and fermionic cases separately. In this section, our calculations primarily follow the method in our previous paper \cite{He:2024xjp}, which involves expanding the Wightman power spectrum into a series form and keeping only the terms that play a major role in the results. However, as $\mu > 0$, the integration will range from $-\infty$ to $\mu$, which means the integration region passes through the origin, making the method of \cite{He:2024xjp} ineffective. Therefore, this paper only considers the case with $\mu \leq 0$.  

{ This section presents our main computational results. We separately analyze bosonic and fermionic Schr\"odinger fields in the grand canonical ensemble. For each case, we: 
	\begin{itemize}
		\item  derive the Wightman power spectrum from the spectral function;
		\item  compute moments using the power spectrum;
		\item  determine Lanczos coefficients $\{a_{n}\},\{b_{n}\}$ via moment method;
		\item  solve the discrete Schr\"odinger equation to obtain the Krylov complexity.
	\end{itemize} 
	Key findings include the linear growth of the Lanczos coefficients and the universal exponential scaling of complexity at late times.}

\subsection{Bosonic case} 
{ We first consider the bosonic field and derive the Wightman power spectrum with the chemical potential $\mu = 0$. The Lanczos coefficients in this case exhibit linear growth with slopes independent of $\mu$. For $\mu < 0$, we demonstrate that while $\{b_{n}\}$ remain $\mu$-independent, $\{a_{n}\}$ acquire a $\mu$-dependent offset. The Krylov complexity grows exponentially at late times.}
\subsubsection{$\mu=0$}
When the chemical potential vanishes, i.e., $\mu = 0$, the Wightman power spectrum of bosonic field $f^W(\omega)$ is 
\begin{equation}\label{fWbosonic}
	f^{W}(\omega)=\mathcal{N}(0,\beta,5,1)\frac{\omega}{2\sinh(\frac{\beta \omega}{2})}\Theta(-\omega).
\end{equation}
According to the normalization condition \eqref{2.31}, the normalization coefficient becomes
\begin{equation}
	\mathcal{N}(0,\beta,5,1)=\frac{4\beta^{2}}{\pi}.
\end{equation}
Subsequently, from the definition of the moment \eqref{2.30}, it is obtained that
\begin{equation}
	\mu_{n}=\frac{2 (-1)^n \left(2^{n+2}-1\right) \beta ^{-n} \zeta (n+2) \Gamma (n+2)}{\pi ^2},
\end{equation}
where $\zeta(s)=\sum_{n=1}^{\infty}\frac{1}{n^{s}}$ is the Riemann zeta function, and $\Gamma(z)=\int_{0}^{\infty}t^{z-1}e^{-t}dt$ is the Gamma function.  

\begin{figure}[htb]
	\centering
\subfigure[]{\includegraphics[width=0.45\textwidth]{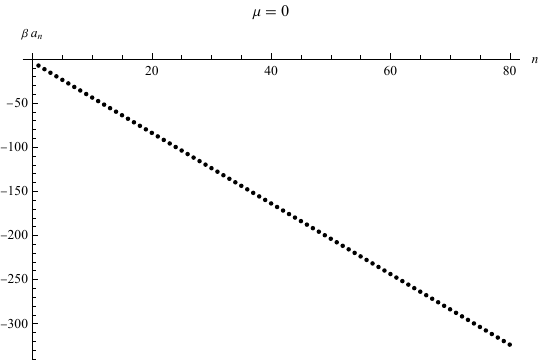}}
\hspace{0.05\textwidth}
\subfigure[]{\includegraphics[width=0.45\textwidth]{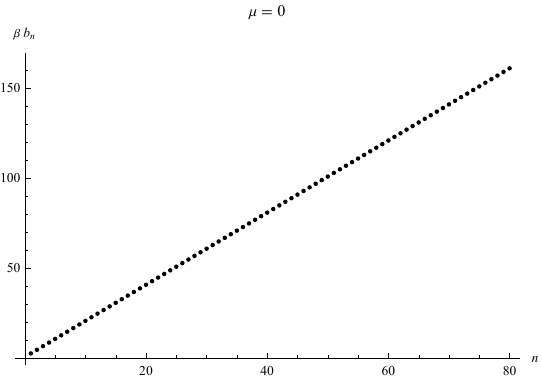}}
\caption{(a) Lanczos coefficients $a_n$ in the bosonic case as $\mu = 0$; (b) Lanczos coefficients $b_n$ in the bosonic case as $\mu = 0$.}
\label{lanczos1}
\end{figure}

\begin{figure}
 	\centering
 	\includegraphics[width=0.6\linewidth]{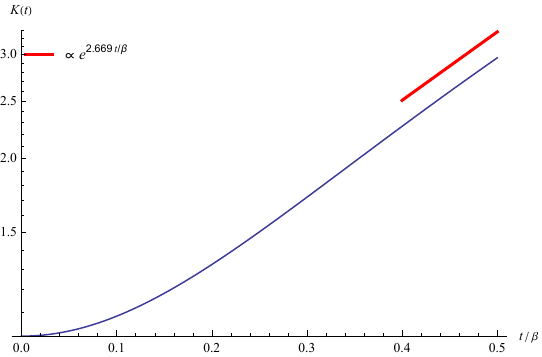}
 	\caption{Time evolution of the Krylov complexity $K(t)$ in the bosonic case as $\mu = 0$. The vertical axis is on a logarithmic scale. The red line indicates the asymptotic behavior of $K(t)$. 	}
 	\label{fig:kt1}
 \end{figure}

Consequently, we can compute the Lanczos coefficients from the recurrence relations in Section \ref{sec2.1}. We present the numerical results of the Lanczos coefficients $\{a_n\}$ and $\{b_n\}$ in Figure \ref{lanczos1}.
% TODO: \usepackage{graphicx} required
It is found that the sequences $\{a_n\}$ and $\{b_n\}$ are almost straight lines with respect to $n$. We obtain the following results after fitting the data in Figure \ref{lanczos1},
\begin{eqnarray}
	\beta a_{n}&\approx&-4.00n-3.96,\label{fitboso1}\\
	\beta b_{n}&\approx&2.00 n+0.98.\label{fitboso2}
\end{eqnarray} 
From equations \eqref{2.26} and \eqref{2.29}, it is known that
\begin{equation}
\begin{aligned}
		\varphi_{0}(t)&=\frac{1}{2\pi}\int_{-\infty}^{\infty}f^{W}(\omega)e^{i\omega t}d\omega\\
		&=\frac{2 \psi ^{(1)}\left(\frac{i t}{\beta }+\frac{1}{2}\right)}{\pi ^2},
\end{aligned}
\end{equation}
in which $f^W$ is from Eq.\eqref{fWbosonic} and $\psi^{(1)}(z)=\frac{d\psi(z)}{dz}$ with $\psi(z)$ the digamma function
\begin{equation}
	\psi(z)=\frac{d}{dz}\ln\Gamma(z)=\frac{\Gamma^{\prime}(z)}{\Gamma(z)}.
\end{equation}
 According to the Schr\"odinger equation \eqref{2.23}, the numerical results of $\varphi_n(t)$ can be readily obtained from the Lanczos coefficients and $\varphi_0(t)$. We must keep the normalization condition \eqref{2.21} of $\varphi_n(t) $ during the numerical computation. 
 % TODO: \usepackage{graphicx} required
 \begin{figure}[htb]
 	\centering
 	\includegraphics[width=0.6\linewidth]{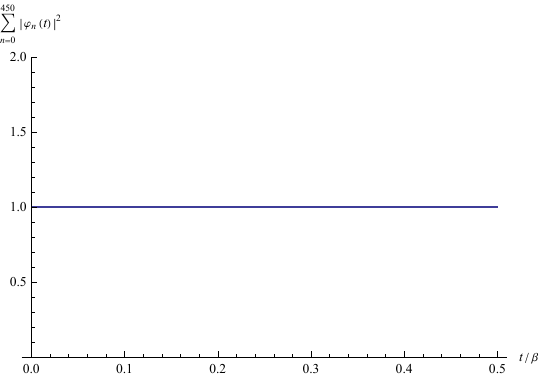}
 	\caption{The normalization condition of $\varphi_n(t)$ is satisfied during the time evolution.}
 	\label{fig:normal1}
 \end{figure}
 Figure \ref{fig:normal1} shows that $\varphi_n(t)$ are indeed normalized during the time evolution, which guarantees that our numerical computation of $\varphi_n(t)$ is correct. In the calculations, we take the maximum value of $n$ to be $450$, which is sufficiently large within the time range we are considering to get the correct values of $\varphi_n(t)$ and the subsequent Krylov complexity. We must stress that the normalization condition of $\varphi_n(t)$ is also well satisfied in other cases in the following; therefore, we do not intend to show them again. In Figure \ref{fig:kt1}, we present the numerical results of the Krylov complexity against the time (scaled by $\beta$), in which the vertical axis is in the logarithmic scale. Therefore, as expected, the Krylov complexity exhibits exponential growth behavior in time. The fitted exponential scaling is roughly $\lambda_K\approx2.669/\beta$, which is smaller than the twice the slope of $\{b_{n}\}$ (which is $4/\beta$). This differs from that in the relativistic free scalar field theory \cite{He:2024xjp}.  This discrepancy is because another Lanczos coefficient  $\{ a_n \}$ will affect the Krylov complexity. It is noted that the computation of complexity as $a_n\neq0$ is mathematically the same as the usual computation of the spread complexity. In \cite{Balasubramanian:2022tpr}, the authors found that when the Hamiltonian can be described as a generator of a certain Lie algebra, the spread complexity can be calculated using the method of generalized coherent states. They consider an example where the Hamiltonian is a generator of SL(2,R)
 	\begin{equation}
 		H=\alpha(L_{-1}+L_{1})+\gamma L_{0}+\delta I,
 	\end{equation}
where $L_{+}$ and $L_{-}$ are raising and lowering ladder operators, and $L_{0}$ belongs to the Cartan subalgebra of the Lie algebra. Then
 \begin{equation}\label{ab2}
 	a_{n}=\gamma(h+n)+\delta,\qquad b_{n}=\alpha\sqrt{n(2h+n-1)},
 \end{equation}
 and the complexity\footnote{Note that the definition of Krylov complexity in \cite{Balasubramanian:2022tpr} differs from ours Eq.\eqref{2.24} by $1$.} is
 \begin{equation}
 	K(t)=\frac{2h}{1-\frac{\gamma^{2}}{4\alpha^{2}}}\sinh^{2}\left (\alpha t\sqrt{1-\frac{\gamma^{2}}{4\alpha^{2}}}\right ).
 \end{equation}
 One can verify that when $a_n = 0$, $K(t)$ will gradually approach a function proportional to $e^{2\alpha t}$ in late time; And when $a_n \neq 0$, it will approach a value proportional to $e^{2\alpha\sqrt{1-\frac{\gamma^{2}}{4\alpha^{2}}}t}$, where $\gamma$ (the scaling of $\{a_n\}$) contributes to the exponential scaling. From the behavior of $\{a_n\}$ and $\{b_n\}$, it is impossible to directly discuss the complexity of the Schr\"odinger field theory using the SL(2,R) Lie algebra. However, we believe that in the Schr\"odinger field theory, the non-Hermitian nature of the field operators leads to the existence of $a_n$, and then this $a_n$ will correct the asymptotic behavior of $K(t)$ like the cases in \cite{Balasubramanian:2022tpr}. This is because the linear relation (by setting $h=1/2$) of $\{a_n\}$ and $\{b_n\}$ in Eq.\eqref{ab2} are similar to those in our linear case Eqs.\eqref{fitboso1} and \eqref{fitboso2} if you are aware of the fact that the Krylov complexity depends much on the behaviors of the Lanczos coefficients. 
  
 % TODO: \usepackage{graphicx} required
 
 \subsubsection{$\mu<0$}
 As chemical potential $\mu < 0$, the Wightman power spectrum $f^W(\omega)$ becomes 
 \begin{equation}\label{4.7}
\begin{aligned}
	 	f^{W}(\omega)&=\mathcal{N}(\mu,\beta,5,1)\frac{(\omega-\mu)}{2}\frac{1}{\sinh\frac{\beta\omega}{2}}\Theta(\mu-\omega)\\
	 	&=\begin{aligned}
	 		\mathcal{N}(\mu,\beta,5,1)(\mu-\omega)\sum_{k=0}^{\infty}e^{\beta\omega(k+1/2)},
	 	\end{aligned}
\end{aligned}
 \end{equation}
 where we have used the equation
 \begin{equation}
 	\frac{1}{1-x}=\sum_{k=0}^{\infty}x^{k},\qquad \text{for $\abs{x}<1$.}
 \end{equation}
 From \eqref{4.7} and \eqref{2.31}, it is known that
 \begin{equation}
 	\mathcal{N}(\mu,\beta,5,1)=\frac{\pi  \beta^2 e^{-\frac{\beta
 				\mu}{2}}}{2 \,
 		_3F_2\left(\frac{1}{2},\frac{1}{2},1;\frac{3}{2},\frac{3}{2}
 		;e^{\beta \mu}\right)},
 \end{equation}
 In which $\,_pF_q$ is the hypergeometric function, defined as
 \begin{equation}
 	\,_pF_q(a_{1},\cdots,a_{p};b_{1},\cdots,b_{q};z)=\sum_{n=0}^{\infty}\frac{(a_{1})_{n}\cdots(a_{p})_{n}}{(b_{1})_{n}\cdots (b_{q})_{n}}\frac{z^{n}}{n!}.
 \end{equation}
 The moments then can be calculated as 
 \begin{equation}
  \begin{aligned}
 	\mu_{n}=&\sum_{k=0}^{\infty}\frac{\beta ^2 (-1)^n 2^{n-1} e^{-\frac{\beta  \mu }{2}} (2 \beta  k+\beta )^{-n-2}}{\,
 		_3F_2\left(\frac{1}{2},\frac{1}{2},1;\frac{3}{2},\frac{3}{2};e^{\beta  \mu }\right)}\\
 		&\times\left [\beta  (2 k+1) \mu  \Gamma \left(n+1,-\frac{1}{2} \beta  (2 k+1) \mu \right)+2 \Gamma \left(n+2,-\frac{1}{2} \beta  (2 k+1) \mu \right)\right ].
 \end{aligned}
 \end{equation}
 It should be noted that, whether for the bosonic case or the fermionic case, in our subsequent numerical calculations, the upper limit of this summation is always taken as $k_{\text{max}} = 200$.
 
\begin{figure}[htb]
 	\centering
 	\subfigure[]{\includegraphics[width=0.45\textwidth]{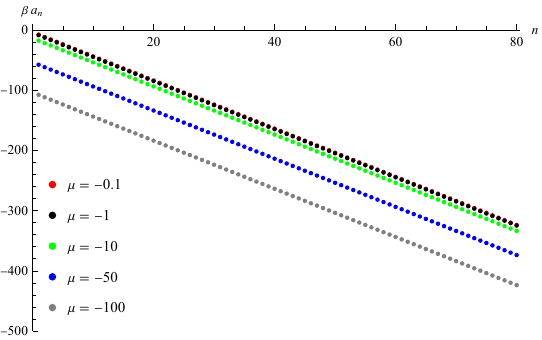}}
 	\hspace{0.05\textwidth}
 	\subfigure[]{\includegraphics[width=0.45\textwidth]{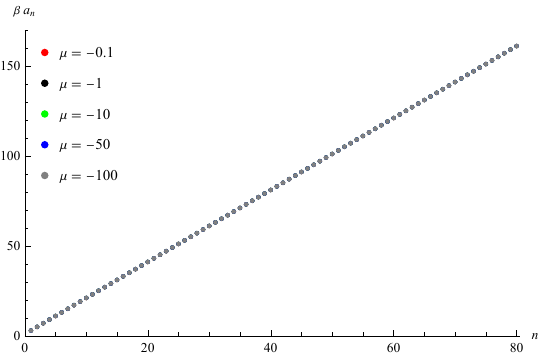}}
 	\caption{(a) Lanczos coefficients $a_n$ in the bosonic case as $\mu <0$ with various chemical potentials; (b) Lanczos coefficients $b_n$ in the bosonic case as $\mu <0$ with various chemical potentials. They overlap together.}
 	\label{lanczos2}
 \end{figure}

  The numerical results of the Lanczos coefficients $a_n$ and $b_n$ are presented in Figure \ref{lanczos2} with various chemical potentials. Interestingly, the behaviors of $b_n$ seem unaffected by the chemical potentials, as seen in panel (b) of Figure \ref{lanczos2}, that they all overlap for various $\mu$'s.
We fit the linear behavior of the $a_n$ and $b_n$ against $n$, and the data are shown in Table \ref{table1}. From the table, we can empirically obtain that the linear fit roughly satisfies  
\begin{equation}\label{4.16}
	{\beta a_{n}\approx-4(n+1)-\mu, \qquad \beta b_n\approx2n+1. }
\end{equation}

\begin{figure}[htb]
	\centering
	\includegraphics[width=0.6\linewidth]{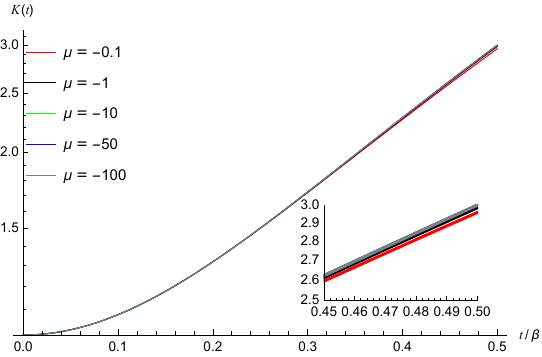}
	\caption{Time evolution of the Krylov complexity in the bosonic case with various chemical potentials. They almost overlap together. The vertical axis is on a logarithmic scale. The inset plot enlarges the figure in the $t/\beta\in[0.45, 0.5]$ range. }
	\label{fig:kt2}
\end{figure}

\begin{table}[h]
	\centering
	\begin{tabular}{|c|c|c|}
	\hline
	Chemical potential& $\beta a_{n}$  &$\beta b_{n}$  \\
	\hline
	$0$& $-3.960875810 - 4.000032607 n$ & $0.9804415923 + 2.000016298 n$ \\
	\hline
	$-0.1$&$-4.075416651 - 4.000020537 n$  &$0.9866122122 + 2.000011603 n$  \\
	\hline
	$-1$&$-4.993572876 - 4.000005364 n$  & $0.9955512958 + 2.000004221 n$ \\
	\hline
	$-10$& $-13.99999932 - 4.000000001 n$ & $0.9987444215 + 2.000001570 n$ \\
	\hline
	$-50$& $-54.00000000 - 4.000000000 n$ & $0.9987447602 + 2.000001569 n$ \\
	\hline
	$-100$& $-104.0000000 - 4.000000000 n$ & $0.9987447602 + 2.000001569 n$ \\
	\hline
\end{tabular}
\caption{Data of the linear fit of the Lanczos coefficients $a_n$ and $b_n$ with respect to $n$, as $\mu \leq 0$ in the bosonic case.}
\label{table1}
\end{table}
 Figure \ref{fig:kt2} shows the time evolution of the Krylov complexity, in which the vertical axis is on a logarithmic scale. We can find that it exponentially increases over a long time.
% TODO: \usepackage{graphicx} required
Besides, as time is short, the Krylov complexities obtained from various chemical potentials almost overlap. However, as time goes on, they start to differ; moreover, the larger the chemical potential, the smaller the Krylov complexity. We have listed the asymptotic slope of $K(t)$ in the Table \ref{t3}. We find that as the chemical potential decreases, the asymptotic slope of $\log K(t)$ increases and approximately approaches  $2.746 (<4)$ (see the second column for bosonic case). 

\begin{table}
	\centering
	\begin{tabular}{|c|c|c|}
		\hline
		{ Chemical potential } &	{ Asymptotic slope of $\log K(t)$} & { Asymptotic slope of $\log K(t)$} \\
		{ $\mu$}  & {(bosonic case in the unit of $1/\beta$)} & {(fermionic case in the unit of $1/\beta$)}\\
		\hline
		$0$ & $2.669$&$2.783 $ \\
		\hline
		$-0.1$ & $2.682$& $2.780$\\
		\hline
		$-1$&$2.726$&$2.761$\\
		\hline
		$-10$&$2.746$ &$2.746$\\
		\hline
		$-50$&$2.746$&$2.746$\\
		\hline
		$-100$&$2.746$&$2.746$\\
		\hline
	\end{tabular}
	\caption{The asymptotic slope of $\log K(t)$ for bosonic and fermionic cases with various chemical potentials.}
	\label{t3}
\end{table}

\subsection{Fermionic case}
For the fermionic case, the Wightman power spectrum differs from the bosonic case \eqref{3.15}, and the operators satisfy the anti-commutative relations. Therefore, we expect that the Krylov complexity in the fermionic case will differ from that in the bosonic case. 
{ The fermionic analysis parallels the bosonic case but with crucial differences in the Wightman power spectrum. Despite the distinct power spectrum, we find remarkably similar Lanczos coefficient structures in fermionic and bosonic cases. Our comparative analysis reveals that while the early-time complexity growth overlaps between bosonic and fermionic cases, the late-time behavior diverges due to the differences in the evolution of $|\varphi_{0}(t)|^{2}$.}
\subsubsection{$\mu=0$}\label{4.2.1}
\begin{figure}[htb]
	\centering
	\subfigure[]{\includegraphics[width=0.45\textwidth]{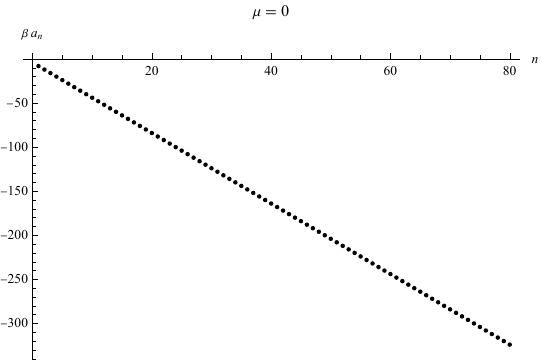}}
	\hspace{0.05\textwidth}
	\subfigure[]{\includegraphics[width=0.45\textwidth]{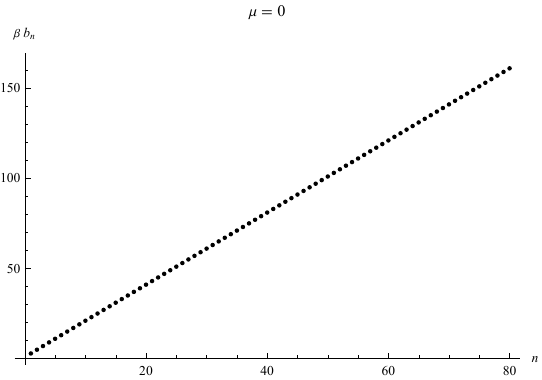}}
	\caption{(a) Lanczos coefficients $a_n$ in the fermionic case as $\mu = 0$; (b) Lanczos coefficients $b_n$ in the fermionic case as $\mu = 0$.}
	\label{lanczos3}
\end{figure}

For $\mu=0$, the Wightman power spectrum $f^{W}(\omega)$ is 
\begin{equation}\label{4.12}
	f^{W}(\omega)=-\mathcal{N}(0,\beta,5,-1)\frac{\omega}{2}\frac{1}{\cosh(\frac{\beta\omega}{2})}\Theta(-\omega),
\end{equation}
where
\begin{equation}
	\mathcal{N}(0,\beta,5,-1)=\frac{\pi  \beta^2}{2 G},\qquad G=\sum_{n=0}^{\infty}\frac{(-1)^{n}}{(2n+1)^{2}}.
\end{equation}
The corresponding moments are
\begin{equation}
	\mu_{n}=\frac{(-1)^n 2^{-n-4} \beta^{-n} \left(\zeta
		\left(n+2,\frac{1}{4}\right)-\zeta
		\left(n+2,\frac{3}{4}\right)\right) \Gamma
		(n+2)}{G}.
\end{equation}
In Figure \ref{lanczos3}, we present the numerical results for the Lanczos coefficients $a_n$ and $b_n$. Surprisingly, these results are similar to the bosonic case with a chemical potential of zero in Figure \ref{lanczos1}.

\begin{figure}[htb]
	\centering
	\subfigure[]{\includegraphics[width=0.45\textwidth]{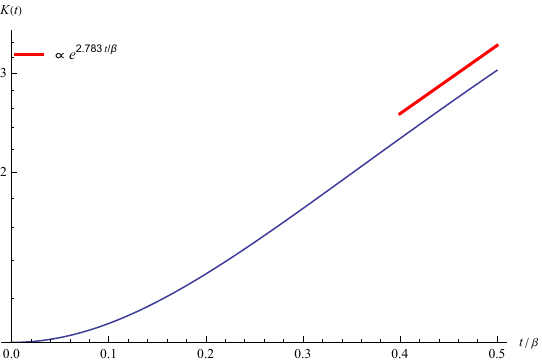}}
	\hspace{0.05\textwidth}
	\subfigure[]{\includegraphics[width=0.45\textwidth]{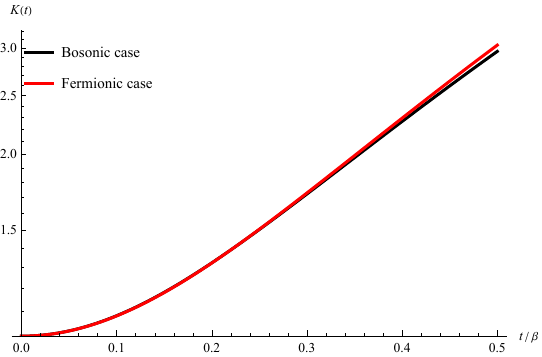}}
	\caption{(a) The Krylov complexity for the fermionic case as $\mu = 0$. The red line indicates the asymptotic behavior of $K(t)$. (b) The comparison of the Krylov complexity for both bosonic case and fermionic cases as $\mu = 0$. At early times, the two complexities overlap; however, as time goes by, the two complexities gradually separate. }
	\label{kt3bvf}
\end{figure}

The linear fitting of the Lanczos coefficients $a_n$ and $b_n$ are 
\begin{eqnarray}
	\beta a_{n}&\approx&-4.00 n-4.01,\label{fitfermi1}\\
	\beta b_{n}&\approx&2.00n+1.00.\label{fitfermi2}
\end{eqnarray}
Interestingly, these fittings for the fermionic case are very close to those in the bosonic case \eqref{fitboso1} and \eqref{fitboso2}.
From \eqref{4.12}, we can get the auto-correlation function as
\begin{equation}\label{4.15}
	\varphi_{0}(t)=\frac{\psi ^{(1)}\left(\frac{it}{2\beta}+\frac{1}{4}\right)-\psi ^{(1)}\left(\frac{i t}{2 \beta }+\frac{3}{4}\right)}{16 G}.
\end{equation}
Based on the Lanczos coefficients and the auto-correlation function \eqref{4.15}, we numerically compute the time evolution of the Krylov complexity; please refer to Figure \ref{kt3bvf}.
Figure \ref{kt3bvf}(a) shows the Krylov complexity for the fermionic case with $\mu = 0$. The vertical axis uses a logarithmic scale. Therefore, we can see that at large time, the linear behavior of $K(t)$ indicates the exponential increase of the Krylov complexity.  As in the bosonic case, the asymptotic behavior of $K(t)$ in the fermionic case is less than twice the slope of $b_n$. As shown in the Figure \ref{kt3bvf}(a), the asymptotic behavior of $K(t)$ is proportional to $e^{2.783t/\beta}$. In Figure \ref{kt3bvf}(b), we compare the Krylov complexities of the bosonic and fermionic cases, and it can be seen that the two curves coincide at early times. However, the two curves will separate as time passes, and the fermionic complexity grows slightly faster than the bosonic one. 

% TODO: \usepackage{graphicx} required

% TODO: \usepackage{graphicx} required
\begin{figure}[htb]
	\centering
	\includegraphics[width=0.6\linewidth]{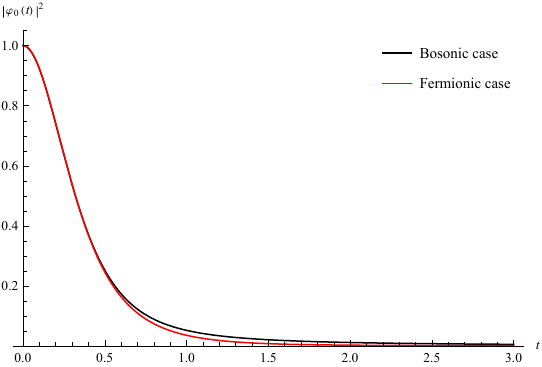}
	\caption{Time evolution of $|\varphi_0(t)|^2$ for the bosonic and fermionic fields as $\mu = 0$. }
	\label{fig:varphiabs}
\end{figure}

\subsubsection{Digression: the similarities of the Krylov complexity for bosonic and fermionic case}\label{similar}
In this part, we will discuss the similarities of the Krylov complexity between bosonic and fermionic cases.  { From the definition of the Krylov complexity in Eq.\eqref{2.24}, we see that the Krylov complexity depends on the wave function $\varphi_n(t)$, more precisely, it directly depends on $\abs{\varphi_n(t)}^{2}$.} Moreover, from the discrete Schr\"odinger equation \eqref{2.23}, it is found that $\varphi_n(t)$ is related to the Lanczos coefficients $a_n$ and $b_n$ as well as the auto-correlation function $\varphi_0(t)$. As we already know, the numerical values of the Lanczos coefficients for the bosonic and fermionic fields are very close to each other for $\mu=0$, see Eqs.\eqref{fitboso1}-\eqref{fitboso2} and Eqs.\eqref{fitfermi1}-\eqref{fitfermi2}. { However, the Wightman power spectrum of bosonic operators is different from that of fermionic operators, which leads to the significant difference in the function $\varphi_{0}(t)$.} { Therefore, we speculate that even if $f^{W}(\omega)$ and $\varphi_{0}(t)$ are different for fermionic and bosonic cases, their $\abs{\varphi_{0}(t)}^{2}$ should exhibit similar behaviors. The small discrepancies in the Krylov complexity between the bosonic and fermionic fields may come from the small differences between $\abs{\varphi_{0}(t)}^{2}$.} We compute and compare the time evolution of $\lvert\varphi_0(t)\lvert^2$ for both bosonic and fermionic fields in Figure \ref{fig:varphiabs}. { From this figure, we can see that the shapes of the two curves are quite similar. At the early time, the curves for $\abs{\varphi_{0}(t)}^{2}$ overlap with each other, but at around $t/\beta\sim0.5$ they begin to separate from each other slightly.} This behavior is consistent with the corresponding time evolution of $K(t)$ in Figure \ref{kt3bvf}(b) that at around $t/\beta\sim0.5$ the Krylov complexity starts to { separate} from each other in numerics.

{ In order to illustrate the behaviors of $\abs{\varphi_{n}(t)}^{2}$ more clearly, we will study the continuous limit of the discrete Schr\"odinger equation in the following. Our conclusion is that if the Lanczos coefficients of the two systems are almost identical and $\abs{\varphi_{0}(t)}^{2}$ are also similar, then $\abs{\varphi_{n}(t)}^{2}$ will not differ from each other significantly.} Therefore, we can assume \cite{Erdmenger:2023wjg,Barbon:2019wsy,Rabinovici:2023yex,Muck:2022xfc}
\begin{eqnarray}
	x_n = \epsilon n, \varphi_n(t)=\varphi(x_n, t), ~{\rm and}~ a_n=a(x_n), b_n = b(x_n), 
\end{eqnarray}
in which $\epsilon$ plays the role of the lattice spacing. That is to say, we will transform the problem of discrete $n$ to the problem of continuous $x_n$ with the small lattice spacing $\epsilon$. Then the discrete Schr\"odinger equation \eqref{2.23} becomes
\begin{equation}
	\partial_{t}\varphi(x_{n},t)=ia(x_{n})\varphi(x_{n},t)+b(x_{n})\varphi(x_{n}-\epsilon,t)-b(x_{n}+\epsilon)\varphi(x_{n}+\epsilon,t).
\end{equation}
We can multiply $\varphi^\ast(x_n,t)$ to both sides of the above equation,
\begin{equation}
	\varphi^{\ast}(x_{n},t)\partial_{t}\varphi(x_{n},t)=ia(x_{n})\abs{\varphi(x_{n},t)}^{2}+b(x_{n})\varphi(x_{n}-\epsilon,t)\varphi^{\ast}(x_{n},t)-b(x_{n}+\epsilon)\varphi(x_{n}+\epsilon,t)\varphi^{\ast}(x_{n},t).
\end{equation}
Make the complex conjugate of the above equation and add them together, then the left-hand side becomes
\begin{equation}
	\text{L.H.S.}=	\varphi^{\ast}(x,t)\partial_{t}\varphi(x,t)+\varphi(x,t)\partial_{t}\varphi^{\ast}(x,t)=\partial_{t}\abs{\varphi(x_{n},t)}^{2}.
\end{equation}
The right-hand side becomes (expand to the order of $\epsilon$)
\begin{equation}
	\begin{aligned}
		\text{R.H.S.}=& b(x_{n})\left[\varphi(x_{n}-\epsilon,t)\varphi^{\ast}(x_{n},t)+\varphi^{\ast}(x_{n}-\epsilon,t)\varphi(x_{n},t)\right]-\\
		& b(x_{n}+\epsilon)\left[\varphi(x_{n}+\epsilon,t)\varphi^{\ast}(x_{n},t)+\varphi^{\ast}(x_{n}+\epsilon,t)\varphi(x_{n},t)\right]\\
		\approx& b(x_{n})[2\abs{\varphi(x_{n},t)}^{2}-\epsilon\partial_{x}\abs{\varphi(x_{n},t)}^{2}]-b(x_{n}+\epsilon)[2\abs{\varphi(x_{n},t)}^{2}+\epsilon\partial_{x}\abs{\varphi(x_{n},t)}^{2}]\\
		=& 2\abs{\varphi(x_{n},t)}^{2}[b(x_{n})-b(x_{n}+\epsilon)]-\epsilon\partial_{x}\abs{\varphi(x_{n},t)}^{2}[b(x_{n})+b(x_{n}+\epsilon)]\\
		\approx &-2\epsilon\abs{\varphi(x_{n},t)}^{2}\partial_{x}b(x_{n})-2\epsilon \partial_{x}\abs{\varphi(x_{n},t)}^{2}b(x_{n}).
	\end{aligned}
\end{equation}
Thus, we obtain the following equation (where we have set $x_n\equiv x$ and $\Psi(x,t)\equiv\abs{\varphi(x,t)}^{2}$),
\begin{equation}\label{Psi}
	\partial_{t}\Psi(x,t)=-2\epsilon\Psi(x,t)\partial_{x}b(x)-2\epsilon\partial_{x}\Psi(x,t)b(x)=-2\epsilon\partial_x\left[\Psi(x,t)b(x)\right].
\end{equation}
This is a partial differential equation of $|\varphi(x,t)|^2$ and $b_n$. We can solve the above equation \eqref{Psi} on an integral curve in the following.  Assuming there is a smooth vector field $V=(\frac{dx}{ds},\frac{dt}{ds})$ on the $t-x$ plane, the integral curves of this vector field are parameterized by $s$, satisfying
\begin{equation}
	\frac{dt}{ds}=1,\qquad \frac{dx}{ds}=2\epsilon b(x).
\end{equation}
Therefore, on this integral curve, we have
\begin{equation}
	\frac{dx}{dt}=2\epsilon b(x).
\end{equation}
Assuming again that this integral curve passes through the point ($x = 0, t = \tau$), then this curve is given by
\begin{equation}
	\int_{0}^{x}\frac{dx^{\prime}}{2\epsilon b(x^{\prime})}=t-\tau.
\end{equation}
in which $\tau$ can be used to label different integral curves. On this single integral curve, we have
\begin{equation}
	\begin{aligned}
		\frac{d}{dt}\Psi(x,t)&=\partial_{t}\Psi(x,t)+\partial_{x}\Psi(x,t)\frac{dx}{dt}\\
		&=-2\epsilon\Psi(x,t)\partial_{x}b(x).
	\end{aligned}
\end{equation}
The solution to the above equation is
\begin{equation}
	\Psi(x,t)=\Psi(0,\tau)\frac{b(0)}{b(x)}=\Psi\left (0,t-\int_{0}^{x}\frac{dx^{\prime}}{2\epsilon b(x^{\prime})}\right )\frac{b(0)}{b(x)}.
\end{equation}
Therefore, as long as we know $\varphi(0,t)$ (equivalent to know $\Psi(0,t)$) and $b(x)$, we can obtain $\Psi(x,t)$ in the late time. This is also the reason that as $\mu=0$, the Krylov complexities for bosonic and fermionic fields have similar behaviors since, in that case, $|\varphi_0(t)|^{2}$ and $b_n$ also behave similarly.

\begin{figure}[h]
	\centering
	\includegraphics[width=0.6\linewidth]{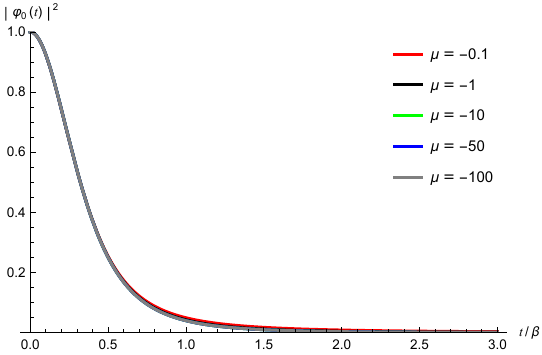}
	\caption{The numerical results of $|\varphi_0(t)|^2$ in the bosonic case with various chemical potentials as $\mu < 0$. The curves are very close to each other.}
	\label{fig:bosephi}
\end{figure}

Returning to the bosonic case, we know that the Krylov complexity for different values of $\mu<0$ are very close to each other (refer to Figure \ref{fig:kt2}) and $b_n$'s are also very close to each other (refer to Table \ref{table1}). Therefore, we expect their values of $\abs{\varphi_0(t)}^2$ to be similar. We present the numerical results for $\abs{\varphi_0(t)}^2$ in the bosonic case with $\mu<0$ in Figure \ref{fig:bosephi}, and it can be found that these curves are indeed very close to each other.

% TODO: \usepackage{graphicx} required

\subsubsection{$\mu<0$}
\begin{figure}[htb]
	\centering
	\subfigure[]{\includegraphics[width=0.45\textwidth]{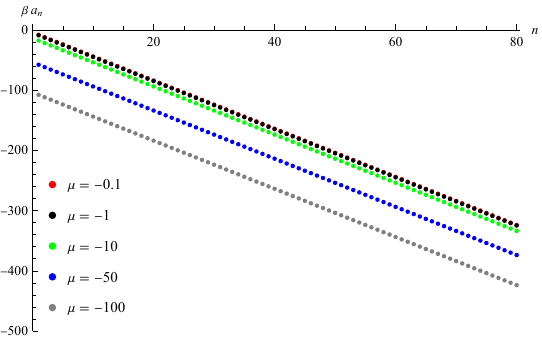}}
	\hspace{0.05\textwidth}
	\subfigure[]{\includegraphics[width=0.45\textwidth]{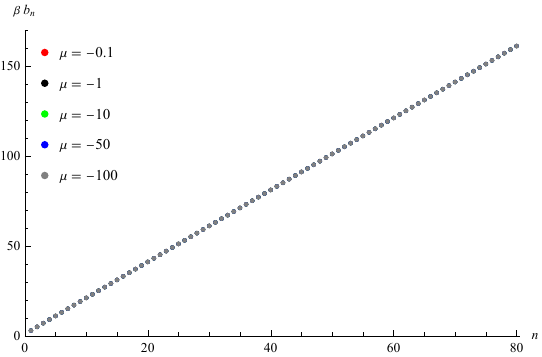}}
	\caption{(a) Lanczos coefficients $a_n$ in the fermionic case as $\mu <0$; (b) Lanczos coefficients $b_n$ in the fermionic case as $\mu <0$. They overlap together for various chemical potentials.}
	\label{lanczos4}
\end{figure}

Let us turn back to the fermionic case with the chemical potential $\mu<0$, the Wightman power spectrum is
\begin{equation}\label{4.34}
\begin{aligned}
		f^{W}(\omega)&=\mathcal{N}(\mu,\beta,5,-1)\frac{\mu-\omega}{2}\frac{1}{\cosh\frac{\beta\omega}{2}}\Theta(\mu-\omega)\\
		&=\mathcal{N}(\mu,\beta,5,-1)(\mu-\omega)\frac{e^{\beta\omega/2}}{1+e^{\beta \omega}}\Theta(\mu-\omega)\\
		&=\mathcal{N}(\mu,\beta,5,-1)(\mu-\omega)\sum_{k=0}^{\infty}(-1)^{k}e^{\beta\omega(k+1/2)}\Theta(\mu-\omega).
\end{aligned}
\end{equation}
From the normalization condition \eqref{2.31}, the constant in the above equation \eqref{4.34} can be determined as
\begin{equation}
	\mathcal{N}(\mu,\beta,5,-1)=\frac{\pi  \beta ^2 e^{-\frac{\beta  \mu }{2}}}{2 \, _3F_2\left(\frac{1}{2},\frac{1}{2},1;\frac{3}{2},\frac{3}{2};-e^{\beta  \mu }\right)}.
\end{equation}
Consequently, the moments $\mu_n$ becomes
\begin{equation}
	\begin{aligned}
		\mu_{n}=&\sum _{k=0}^{\infty } \frac{\beta ^2 2^{n-1} e^{-\frac{\beta  \mu }{2}} (-1)^{k+n} (2 \beta  k+\beta )^{-n-2}}{\,
		_3F_2\left(\frac{1}{2},\frac{1}{2},1;\frac{3}{2},\frac{3}{2};-e^{\beta  \mu }\right)}\\
		 &{\times}\left(\beta  (2 k+1) \mu  \Gamma
		\left(n+1,-\frac{1}{2} \beta  (2 k+1) \mu \right)+2 \Gamma \left(n+2,-\frac{1}{2} \beta  (2 k+1) \mu \right)\right).
	\end{aligned}
\end{equation}
 The resulting Lanczos coefficients $a_n$ and $b_n$ are exhibited in Figure \ref{lanczos4}.

\begin{table}[h]
	\centering
	\begin{tabular}{|c|c|c|}
		\hline
		Chemical potential& $\beta a_{n}$  &$\beta b_{n}$  \\
		\hline
		$0$& $-4.011472827 - 3.999990429 n$ & $1.004458037 + 1.999996819 n$ \\
		\hline
		$-0.1$&$-4.110594174 - 3.999991162 n$  &$1.004020088 + 1.999997183 n$  \\
		\hline
		$-1$&$-5.004915662 - 3.999995899 n$  & $1.001191152 + 1.999999536 n$ \\
		\hline
		$-10$& $-14.00000068 - 3.999999999 n$ & $0.9987450989 + 2.000001569 n$ \\
		\hline
		$-50$& $-54.00000000 - 4.000000000 n$ & $0.9987447602 + 2.000001569 n$ \\
		\hline
		$-100$& $-104.0000000 - 4.000000000 n$ & $0.9987447602 + 2.000001569 n$ \\
		\hline
	\end{tabular}
	\caption{The numerical results of Lanczos coefficients when $\mu \leq 0$ in the fermionic case.} 
\label{table2}
\end{table}
We list the numerical results of the fittings in the Table \ref{table2} for the fermionic case. Similar to the bosonic case, the coefficients $a_n$ are almost parallel for various chemical potentials, while $b_n$ almost overlap. From the Table \ref{table2}, the linear relations are roughly, 
\begin{eqnarray}
	\beta a_{n}&\approx&-4(n+1)-\mu,\\
	\beta b_{n}&\approx&2n+1.
\end{eqnarray}

\begin{figure}[htb]
	\centering
	\subfigure[]{\includegraphics[width=0.45\textwidth]{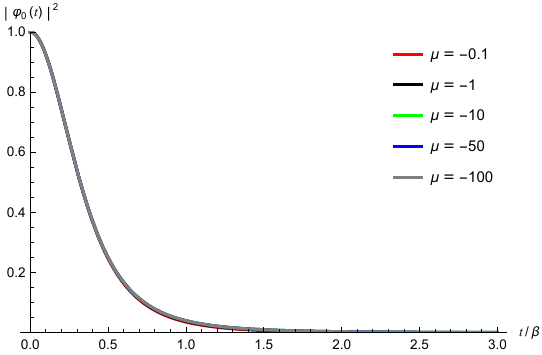}}
	\hspace{0.05\textwidth}
	\subfigure[]{\includegraphics[width=0.45\textwidth]{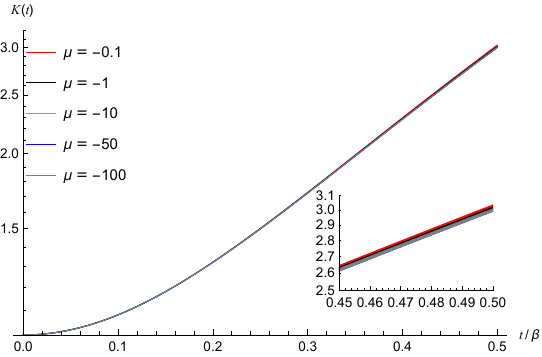}}
	\caption{(a) The profile of $|\varphi_0(t)|^2$ for the fermionic case at different chemical potentials when $\mu < 0$; (b) The Krylov complexity for the fermionic case at different chemical potentials when $\mu < 0$.}
	\label{fig:kt4}
\end{figure}
The corresponding $\lvert\varphi_0(t)\rvert^2$ of the auto-correlation functions and the Krylov complexity are shown in Figure \ref{fig:kt4}. It is found that $\lvert\varphi_0(t)\rvert^2$ almost overlap with the Krylov complexities for various chemical potentials. This is consistent with our discussions in the preceding subsection \ref{similar} that if the Lanczos coefficients $b_n$ are similar to each other, and $\lvert\varphi_0(t)\rvert^2$ are also similar, then the Krylov complexities will overlap. However, it is interesting to see that, although the Krylov complexities almost overlap, there are still some minor discrepancies between them. In the inset plot of Figure \ref{fig:kt4}, we can still distinguish some minor discrepancies between the Krylov complexities for various chemical potentials at late time. The inset plot shows that unlike in the bosonic case (see Figure \ref{fig:kt2}), the larger the chemical potential, the faster the Krylov complexity grows. Table \ref{t3} presents the asymptotic slope of $\log K(t)$ for the fermionic case. Unlike the bosonic case, it is interesting to see that in the fermionic case, the smaller the chemical potential is, the smaller the slope will be. Finally, they will approach $2.746$, the same as the bosonic case.

\section{Conclusions}\label{sec5}
In this paper, we have applied the Krylov subspace method in the grand canonical ensemble to investigate the Krylov complexity of the non-relativistic Schr\"odinger field with various chemical potentials, especially as the chemical potential is zero or negative.  
We found that regardless of whether the chemical potential is zero or negative, or whether it is for bosonic or fermionic case, the Lanczos coefficients $\{a_n\}$ and $\{b_n\}$ are approximately linear and very close to
\begin{eqnarray*}
	\beta a_{n}&\approx&-4(n+1)-\mu,\\
	\beta b_{n}&\approx& 2n+1.
\end{eqnarray*}
This is somewhat unexpected, as the Wightman power spectrum and the auto-correlation functions differ in various situations. However, they possess similar Lanczos coefficients and consequently exhibit similar behaviors of Krylov complexity. We suspect that there must be some key quantities that are similar between them, and we find that this quantity is the square of the absolute values of the auto-correlation function $\lvert\varphi_0(t)\rvert^2$. That is to say, similar behavior of the $\lvert\varphi_0(t)\rvert^2$ for bosonic and fermionic cases will yield similar behaviors of the Krylov complexity. { It is worth noting that our Lanczos coefficients always absent staggering, which is different from some previous results in field theory. Therefore, by examining examples in Appendix \ref{appc}, we found that the conditions for the staggering presented in \cite{Camargo:2022rnt} can be further generalized. Our conclusion is
	\begin{itemize}
	\item {\it if the Wightman power spectrum is symmetric and the zero point is located at the symmetry axis, the Lanczos coefficients $b_n$ will exhibit staggering. }
	\item {\it if the Wightman power spectrum is asymmetric, the Lanczos coefficients $b_n$ will not be staggering.}
	\end{itemize}
}

However, in detail, there are still some discrepancies between the bosonic and fermionic cases. We found that the Krylov complexities for both bosonic and fermionic cases will overlap at early times. However, as time passes, the Krylov complexity for the bosonic case with a larger chemical potential grows slightly faster than that with a smaller chemical potential. On the contrary, for the fermionic case, the Krylov complexity with a smaller chemical potential grows slightly faster than that with a larger chemical potential as time increases. 
By fitting the slopes of $K(t)$ at a later time, we found that for the bosonic case, as the chemical potential decreases, the asymptotic slope increases. However, for the fermionic case, as the chemical potential decreases, the asymptotic slope decreases. { It is difficult to explain the discrepancies of the Krylov complexity for the bosonic and fermionic cases precisely, but we can speculate that these discrepancies must be related to the different statistical properties of the bosonic and fermionic fields, or equivalently the algebraic structure for them. The Bose-Einstein distribution for the bosonic field is { $n_{B}(\omega)=1/(e^{-\beta\omega}-1)$}\footnote{For free bosons, the Bose distribution usually given in textbooks is $1/(e^{(E_{\mathbf{k}}-\mu)\beta}-1)$. From the first line of Eq. (3.13) in the manuscript, we know $E_{\mathbf{k}}-\mu=-\omega$, so the Bose distribution can be written as $n_{B}(\omega)=1/(e^{-\beta\omega}-1)$. For free fermions, we get $n_{F}(\omega)=1/(e^{-\beta\omega}+1)$.} and the Fermi-Dirac distribution for the fermionic field is {$n_{F}(\omega)=1/(e^{-\beta\omega}+1)$}. Fortunately, we found that the Wightman power spectrum \eqref{3.15} can be rewritten as 
	{	\begin{equation}\label{d1}
			f^{W}(\omega)=\left \{\begin{aligned}
				&\mathcal{N}(\mu,\beta,5,1)(\mu-\omega)e^{\frac{-\beta\omega}{2}}n_{B}(\omega)\Theta(\mu-\omega),\qquad \text{bosonic field}\\
				&\mathcal{N}(\mu,\beta,5,-1)(\mu-\omega)e^{-\frac{\beta\omega}{2}}n_{F}(\omega)\Theta(\mu-\omega).\qquad \text{fermionic field}
			\end{aligned}\right .
		\end{equation}}
From the above formula we can see that the different statistical distributions for the bosonic and fermionic fields have direct impact to the Wightman power spectrum $f^W(\omega)$, therefore, it then further has impacts to the Krylov complexity. However, it is hard to see the relation between the asymptotic scaling of Krylov complexity and the chemical potential, since as we compute the Krylov complexity we need to Fourier transform $f^W(\omega)$ to get the autocorrelation function $\varphi_0(t)$. Then we use the Lanczos coefficients and $\varphi_0(t)$ to get the values of $\varphi_n(t)$. Finally, the Krylov complexity is obtained by summing up the $n\abs{\varphi_n(t)}^2$. These tedious processes make us hard to tell the exact relations between the chemical potential and the Krylov complexity, but rather we can only see the relations from numerical computations.} Moreover, both in the bosonic and fermionic cases, the asymptotic slope will approach $2.746$ for small chemical potentials. {  This is because when the chemical potential is small enough, the power spectra of the Boson field and Fermi field gradually converge.} Interestingly, this slope is smaller than twice the slope of $b_{n}$. This is because we are considering the Krylov complexity of non-Hermitian operators; in this case, the Lanczos coefficients $a_n$ will contribute to the Krylov complexity. We expect that our work will shed light on the study of Krylov complexity of the non-relativistic field theory with non-Hermitian operators.

\acknowledgments
This work was partially supported by the National Natural Science Foundation of China (Grants No.12175008). 

\appendix
\section{Solve the discrete Schr\"odinger equation}\label{appa}
To solve the discrete Schr\"odinger equation, we need first to discretize time, letting $t_i \approx (i-1) \Delta t,$ where $i = 1, 2, \ldots$, $\Delta t$ is a small time interval. Then we have
\begin{equation}\label{A1}
	\begin{aligned}
		\varphi_{n}(t_{i+1})&=\varphi_{n}(t_{i}+\Delta t)\\
		&\approx \varphi_{n}(t_{i})+\partial_{t}\varphi_{n}(t_{i})\Delta t\\
		&=\varphi_{n}(t_{i})+[b_{n}\varphi_{n-1}(t_{i})-b_{n+1}\varphi_{n+1}(t_{i})+ia_{n}\varphi_{n}(t_{i})]\Delta t.
	\end{aligned}
\end{equation}
Write $\varphi_{n}(t_i)$ as $\varphi^i_n$ and introduce the vector $\vec{\varphi}^i = (\varphi^{i}_{0}\ \varphi^{i}_{1}\cdots)^{\text{T}}$, then \eqref{A1} can be written in a more compact form
\begin{equation}\label{A.2}
	\vec{\varphi}^{i+1}=\vec{\varphi}^{i}+A\vec{\varphi}^{i}\Delta t,
\end{equation}
where $A$ is a matrix
\begin{equation}
	A=\left (\begin{matrix}
		ia_{0}&-b_{1}&0&0&0&\cdots\\
		b_{1}&ia_{1}&-b_{2}&0&0&\cdots\\
		0&b_{2}&ia_{2}&-b_{3}&0&\cdots\\
		\vdots&\vdots&\vdots&\vdots&\vdots&\ddots
	\end{matrix}\right ).
\end{equation}
To utilize the fourth-order Runge-Kutta method, \eqref{A.2} can be rewritten as
\begin{equation}\label{A.4}
	\vec{\varphi}^{i+1}=\vec{\varphi}^{i}+\frac{\Delta t}{6}(K_{1}+2K_{2}+2K_{3}+K_{4}),
\end{equation}
where
\begin{gather}
	K_{1}=A\vec{\varphi}^{i},\qquad K_{2}=A(\vec{\varphi}^{i}+K_{1}\Delta t/2),\\
	 K_{3}=A(\vec{\varphi}^{i}+K_{2}\Delta t/2),\qquad K_{4}=A(\vec{\varphi}^{i}+K_{3}\Delta t).
\end{gather}
Once the Lanczos coefficients $\{a_n\}$ and $\{b_n\}$ are obtained, it is equivalent to knowing the matrix $A$. Knowing $\varphi_n(0)$ also means knowing $\vec{\varphi}^1$, and from \eqref{A.4}, one can calculate $\vec{\varphi}^i$.

\section{Derive equation \eqref{3.9}}\label{appB}
Here, only the bosonic case is considered, and the results for the fermionic case are the same as those for the bosonic case. Define the retarded correlator
\begin{equation}
	D^{R}_{ab}(t,\mathbf{x};t^{\prime},\mathbf{x}^{\prime})\equiv i\expval{\Theta(t-t^{\prime})[\phi_{a}(t,\mathbf{x}),\psi_{b}(t^{\prime},\mathbf{x}^{\prime})]}.
\end{equation}
Note that 
\begin{equation}
	\Theta(t)=i\int_{-\infty}^{\infty}\frac{d\omega}{2\pi}\frac{e^{-i\omega t}}{\omega+i\epsilon},\qquad \epsilon=0^{+}.
\end{equation}
Therefore, in momentum space, we have
\begin{equation}
	D^{R}_{ab}(\omega,\mathbf{k})=\int_{-\infty}^{\infty}\frac{d\omega^{\prime}}{2\pi}\frac{\rho_{ab}(\omega^{\prime},\mathbf{k})}{\omega^{\prime}-\omega-i\epsilon}.
\end{equation}
Make use of the identity
\begin{equation}
	\frac{1}{x\pm i\epsilon}=\mathcal{P}\frac{1}{x}\mp i\pi\delta(x),
\end{equation}
we obtain
\begin{equation}
	\rho_{ab}(\omega,\mathbf{k})=2\Im D^{R}_{ab}(\omega,\mathbf{k}),
\end{equation}
where $\mathcal{P}$ denotes the principal value. On the other hand, the relationship between the imaginary-time correlation function, which is also known as the thermal propagator, and the spectral density is
\begin{equation}
	\mathcal{D}_{ab}(\omega_{n},\mathbf{k})=\int_{-\infty}^{\infty}\frac{d\omega^{\prime}}{2\pi}\frac{\rho_{ab}(\omega^{\prime},\mathbf{k})}{\omega^{\prime}+i\omega_{n}}.
\end{equation}
That is to say, by simply replacing $\omega_n$ with $i\omega - \epsilon$, one can obtain the retarded correlation function from the thermal propagator
\begin{equation}
	D^{R}_{ab}(\omega,\mathbf{k})=\mathcal{D}_{ab}(\omega_n\rightarrow i\omega-\epsilon,\mathbf{k}).
\end{equation}
The thermal propagator of Schr\"odinger field theory is given by \eqref{3.8}, then we have
\begin{equation}
\begin{aligned}
		\rho_{ab}(\omega,\mathbf{k})&=2\Im D^{R}_{ab}(\omega,\mathbf{k})\\
		&=2\Im\mathcal{D}(\omega_{n}\rightarrow i\omega-\epsilon,\mathbf{k})\\
		&=2\Im{\frac{1}{i(i\omega-\epsilon)+\frac{\mathbf{k}^{2}}{2m}-\mu}}\\
		&=2\Im{\frac{1}{\frac{\mathbf{k}^{2}}{2m}-\mu-\omega-i\epsilon}}\\
		&=2\Im{\mathcal{P}\left (\frac{\mathbf{k}^{2}}{2m}-\mu-\omega\right )+i\pi\delta\left (\frac{\mathbf{k}^{2}}{2m}-\mu-\omega\right )}\\
		&=2\pi\delta(\xi_{\mathbf{k}}-\omega).
\end{aligned}
\end{equation}
This is the equation \eqref{3.9} in the main text.

  \section{About staggering}\label{appc}
In \cite{Camargo:2022rnt}, the authors have studied various examples and then summarized that if the following two conditions are satisfied, the Lanczos coefficients will not exhibit staggering behavior:
\begin{enumerate}
	\item The power spectrum is finite and positive at $\omega=0$, i.e., $0<f^{W}(0)<\infty$;
	\item The derivative of the power spectrum $f^{W}(\omega)$ is a continuous function of $\omega$ for $-\Lambda<\omega <\Lambda$, where $\Lambda $ is a UV cutoff.
\end{enumerate}
However, we find that these two conclusions are not complete.  For instance in our manuscript, the power spectrum of the fermionic operator is
\[f^W(\omega)=\mathcal{N}(\mu,\beta,5,-1)\frac{(\mu-\omega)}{2}\frac{1}{\cosh\frac{\beta\omega}{2}}\Theta(\mu-\omega).\]
 If we consider the case of $\mu=0$, then the power spectrum satisfies $f^W(0)=0$. But in this case the Lanczos coefficients are not staggering, see the Fig. \ref{lanczos3}. Therefore, our case violates the above condition (a) in \cite{Camargo:2022rnt} where they asserts that $f^W(0)>0$. 
 
%One reason is that the paper arXiv:2212.14702 only studied the case with Lanczos coefficients $\{b_n\}$. There are no $\{a_n\}$ coefficients in their case. In our manuscript, we have both $\{a_n\}$ and $\{b_n\}$ coefficients. Therefore, we expect that the analysis of the staggering behavior of these coefficients will be different from those in the paper arXiv:2212.14702. The other reason is that 

\begin{figure}[htb]
			\centering
			\subfigure[]{\includegraphics[width=0.45\textwidth]{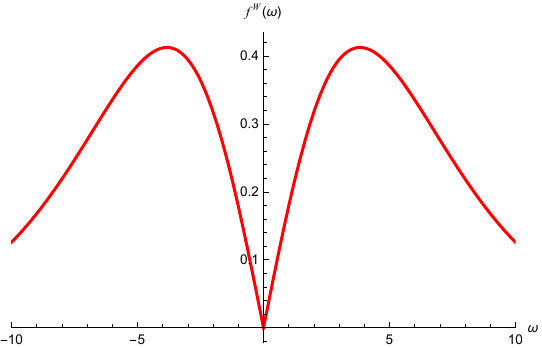}}
			\hspace{0.05\textwidth}
			\subfigure[]{\includegraphics[width=0.45\textwidth]{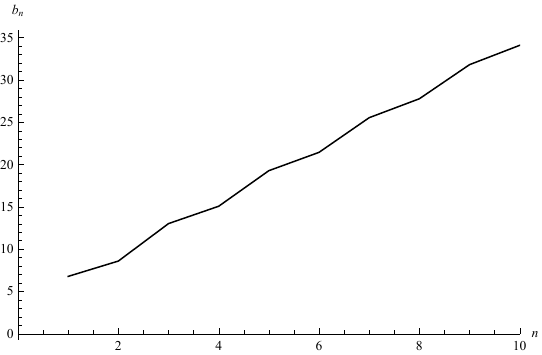}}
			\caption{Figures correspond to example 1. (a) Profile of the power spectrum \eqref{c1}. It is an even function, whose zero point lies at the symmetry axis at $\omega=0$; (b) Lanczos coefficients $b_{n}$ for the symmetric Wightman power spectrum \eqref{c1}. Obviously, $b_{n}$ (points corresponding to integers $n$) exhibits staggering behavior. }
			\label{powerspectrum1}
		\end{figure}

In the following we have found more exceptional examples which do not meet the conclusions in \cite{Camargo:2022rnt}.	First, let's consider the following example %from arXiv:2212.14702:
		\begin{equation}\label{c1}
			(\text{\textbf{example 1}})\qquad 	f^{W}(\omega)=\frac{\pi}{28\zeta(3)}\frac{\omega^{2}}{\sinh(\frac{\abs{\omega}}{2})}.
		\end{equation}
		The Lanczos coefficients of this power spectrum exhibit staggering behavior (see Fig.\ref{powerspectrum1}(b)).
		This result is consistent with the conclusions in \cite{Camargo:2022rnt}. 
		
\begin{figure}[htb]
			\centering
			\subfigure[]{\includegraphics[width=0.45\textwidth]{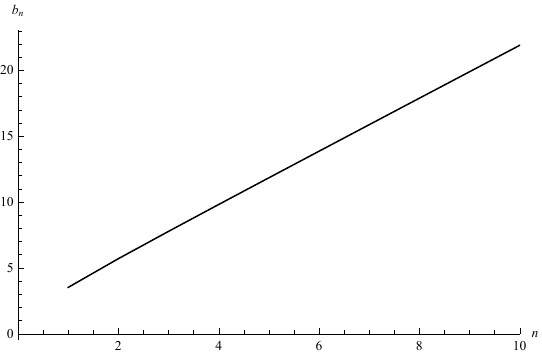}}
			\hspace{0.05\textwidth}
			\subfigure[]{\includegraphics[width=0.45\textwidth]{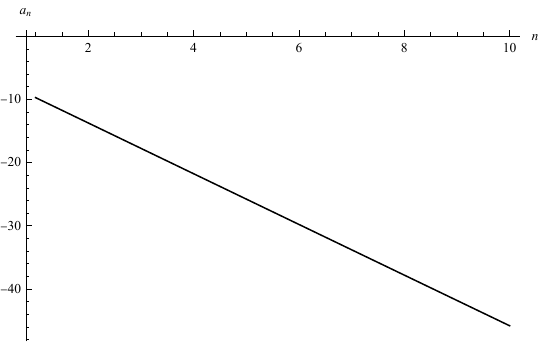}}
			\caption{Figures correspond to example 1'. (a) Lanczos coefficients $b_n$ for power spectrum \eqref{c2}; (b) Lanczos coefficients $a_n$ for power spectrum \eqref{c2}. We can see that $\{a_{n}\}$ and $\{b_{n}\}$ (points corresponding to integer $n$) lie in perfect straight lines.}  %This confirms that staggering requires non-endpoint zero in $f^{W}(\omega)$.}
			\label{powerspectrum2}
		\end{figure}

		However, if we set the part of \eqref{c1} where $\omega \geq0$ to zero,  the Wightman power spectrum \eqref{c1} will change to: \footnote{After setting a certain part of the original power spectrum to zero, the remaining power spectrum needs to be renormalized to satisfy renormalization condition \eqref{2.31}. Therefore, the new normalization constant will be changed.}
		\begin{equation}\label{c2}
			(\text{\textbf{example 1'}})\qquad 	f^{W}(\omega)=\left \{\begin{matrix}
					\frac{\pi}{14\zeta(3)}\frac{\omega^{2}}{\sinh(\frac{\abs{\omega}}{2})},& \omega< 0,\\
					0,& \omega\geq0.
				\end{matrix}
				\right.
			\end{equation}
For the new Wightman power spectrum Eq.\eqref{c2}, we find that the staggering of the Lanczos coefficients disappears, see Fig.\ref{powerspectrum2}.
		There is no staggering behavior of the Lanczos coefficients $\{a_{n}\}$ and $\{b_{n}\}$ in Fig.\ref{powerspectrum2}. But we can see that in this case $f^{W}(0)=0$, which violates the conclusions in \cite{Camargo:2022rnt}! 
		
		By inspecting the paper \cite{Camargo:2022rnt} carefully, we find that all the examples considered in it are even functions which have a symmetry axis at $\omega=0$. Then they can draw the above conclusions. This inspires us to realize that maybe the violation of their conclusions comes from the symmetries of the power spectrum. Specifically, we generalize the conclusion in \cite{Camargo:2022rnt} and conjecture that,
		\begin{itemize}
		\item {\it if the Wightman power spectrum is symmetric and the zero point is located at the symmetry axis, the Lanczos coefficients $b_n$ will exhibit staggering. }
		\item {\it if the Wightman power spectrum is asymmetric, the Lanczos coefficients $b_n$ will not be staggering.}
		\end{itemize}
		
\begin{figure}[htb]
			\centering
			\includegraphics[width=0.6\linewidth]{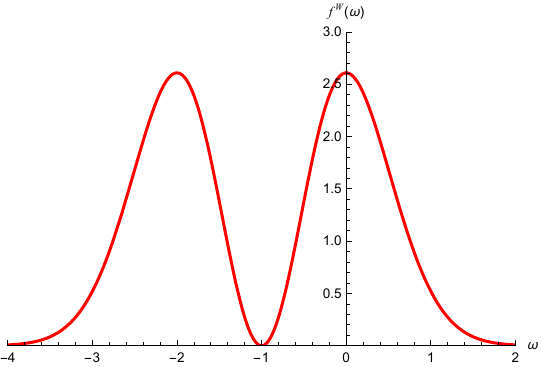}
			\caption{Figure corresponds to example 2. The power spectrum \eqref{c5} is symmetric about the axis $\omega=-1$, with its zeros located at the symmetry axis.}
			\label{fig:power3}
		\end{figure}

\begin{figure}[htb]
			\centering
			\subfigure[]{\includegraphics[width=0.45\textwidth]{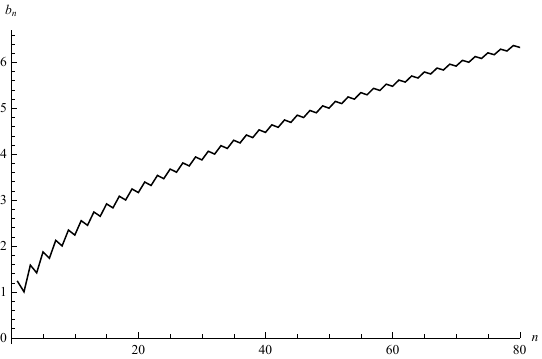}}
			\hspace{0.05\textwidth}
			\subfigure[]{\includegraphics[width=0.45\textwidth]{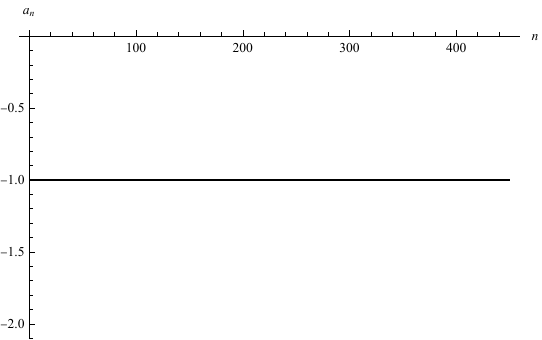}}
			\caption{Figures correspond to example 2. (a) The Lanczos coefficients $b_{n}$ corresponding to the power spectrum \eqref{c5}; (b) The Lanczos coefficients $a_{n}$ corresponding to \eqref{c5}. The coefficients $b_{n}$ exhibit the staggering behavior, while $a_{n}$ remains constant and precisely equals the position of the symmetry axis.}
			\label{powerspectrum6}
		\end{figure}

%========exmaple 2==========		
		To verify this conjecture, we consider the following power spectrum,
		\begin{equation}\label{c5}
			(\text{\textbf{Example} 2})\qquad f^{W}(\omega)=4\sqrt{\pi}(1+\omega)^{2}e^{-(1+\omega)^{2}}.
		\end{equation}
		This power spectrum is symmetric about $\omega=-1$ and $f^{W}(-1)=0$, as shown in Figure \ref{fig:power3}.
		% TODO: \usepackage{graphicx} required
		The Lanczos coefficients of this power spectrum are shown in Figure \ref{powerspectrum6}, where the familiar staggering behavior of $b_{n}$ is observed. This phenomenon violates the conclusions in \cite{Camargo:2022rnt}. But this phenomenon satisfies our conjecture. % is consistent with the arXiv:2212.14702 since the power spectrum \eqref{c5} is symmetric. 
		
		An interesting phenomenon is that now the Lanczos coefficient $a_{n}$ is a constant, i.e., $a_{n}=1$, which equals the value of the symmetry axis. This motivates to examine many examples and finally we find that: {\bf if the Wightman power spectrum is symmetric, then the Lanczos coefficient $a_{n}$ is a constant, equal to the value of the symmetry axis of the power spectrum}.  %when we shift an even function to the left by $1$, the Lanczos coefficient $a_{n}$ change from $a_{n}=0 $ to $a_{n}=−1$. This may suggest that the constant $a_{n} exactly corresponds to the axis of symmetry of $f ^{W}(\omega)$. 
		
\begin{figure}[htb]
			\centering
			\subfigure[]{\includegraphics[width=0.45\textwidth]{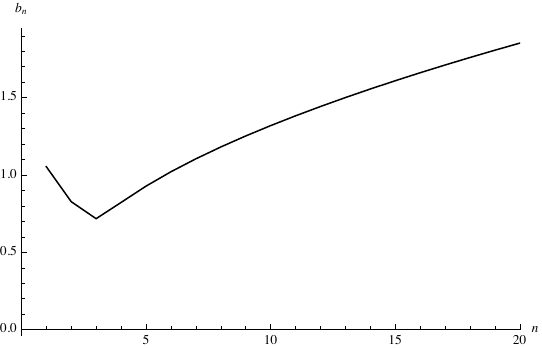}}
			\hspace{0.05\textwidth}
			\subfigure[]{\includegraphics[width=0.45\textwidth]{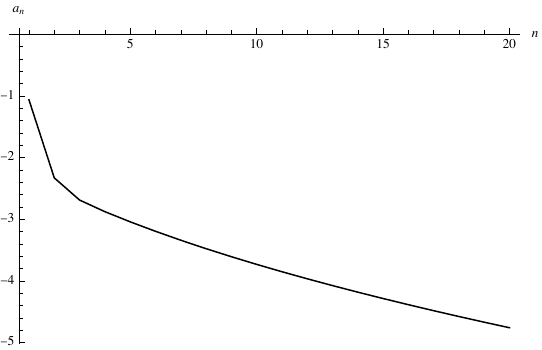}}
			\caption{Figures correspond to example 2'. (a) The Lanczos coefficients $b_{n}$ corresponding to the power spectrum \eqref{c6}; (b) The Lanczos coefficients $a_{n}$ corresponding to the power spectrum \eqref{c6}.  The Lanczos coefficients no longer exhibit staggering behavior, since the power spectrum is asymmetric. }
			\label{powerspectrum7}
		\end{figure}

		Let's consider the asymmetric form of the power spectrum \eqref{c5} by setting the positive frequency part to zero, i.e.,  the new power spectrum becomes:
		\begin{equation}\label{c6}
		(\text{\textbf{example 2'}})\qquad 	f^{W}(\omega)=\left \{\begin{matrix}
				\frac{8\pi}{\sqrt{\pi}(1+\text{Erf}(1))-2e^{-1}}(1+\omega)^{2}e^{-(1+\omega)^{2}},&\qquad&\omega\le0\\
				0,&\qquad&\omega>0,
			\end{matrix}\right .
		\end{equation}
		where Erf($\cdot$) is the error function. Its Lanczos coefficients are shown in Figure \ref{powerspectrum7}. Obviously, we see that the Lanczos coefficients do not exhibit staggering behavior, which supports our conclusions above. 

\color{black}

\newpage
%\bibliographystyle{jhep}
%\bibliography{references.bib}

\begin{thebibliography}{10}
	
	\bibitem{Parker:2018yvk}
	D.E.~Parker, X.~Cao, A.~Avdoshkin, T.~Scaffidi and E.~Altman, \emph{{A
			Universal Operator Growth Hypothesis}},
	\href{https://doi.org/10.1103/PhysRevX.9.041017}{\emph{Phys. Rev. X}
		{\bfseries 9} (2019) 041017}
	[\href{https://arxiv.org/abs/1812.08657}{{\ttfamily 1812.08657}}].
	
	\bibitem{Rabinovici:2022beu}
	E.~Rabinovici, A.~S\'anchez-Garrido, R.~Shir and J.~Sonner, \emph{{Krylov
			complexity from integrability to chaos}},
	\href{https://doi.org/10.1007/JHEP07(2022)151}{\emph{JHEP} {\bfseries 07}
		(2022) 151} [\href{https://arxiv.org/abs/2207.07701}{{\ttfamily
			2207.07701}}].
	
	\bibitem{Liu:2022god}
	C.~Liu, H.~Tang and H.~Zhai, \emph{{Krylov complexity in open quantum
			systems}},
	\href{https://doi.org/10.1103/PhysRevResearch.5.033085}{\emph{Phys. Rev.
			Res.} {\bfseries 5} (2023) 033085}
	[\href{https://arxiv.org/abs/2207.13603}{{\ttfamily 2207.13603}}].
	
	\bibitem{Trigueros:2021rwj}
	F.B.~Trigueros and C.-J.~Lin, \emph{{Krylov complexity of many-body
			localization: Operator localization in Krylov basis}},
	\href{https://doi.org/10.21468/SciPostPhys.13.2.037}{\emph{SciPost Phys.}
		{\bfseries 13} (2022) 037}
	[\href{https://arxiv.org/abs/2112.04722}{{\ttfamily 2112.04722}}].
	
	\bibitem{Bhattacharjee:2022ave}
	B.~Bhattacharjee, P.~Nandy and T.~Pathak, \emph{{Krylov complexity in large q
			and double-scaled SYK model}},
	\href{https://doi.org/10.1007/JHEP08(2023)099}{\emph{JHEP} {\bfseries 08}
		(2023) 099} [\href{https://arxiv.org/abs/2210.02474}{{\ttfamily
			2210.02474}}].
	
	\bibitem{Caputa:2022yju}
	P.~Caputa, N.~Gupta, S.S.~Haque, S.~Liu, J.~Murugan and H.J.R.~Van~Zyl,
	\emph{{Spread complexity and topological transitions in the Kitaev chain}},
	\href{https://doi.org/10.1007/JHEP01(2023)120}{\emph{JHEP} {\bfseries 01}
		(2023) 120} [\href{https://arxiv.org/abs/2208.06311}{{\ttfamily
			2208.06311}}].
	
	\bibitem{Bhattacharjee:2023uwx}
	B.~Bhattacharjee, P.~Nandy and T.~Pathak, \emph{{Operator dynamics in
			Lindbladian SYK: a Krylov complexity perspective}},
	\href{https://doi.org/10.1007/JHEP01(2024)094}{\emph{JHEP} {\bfseries 01}
		(2024) 094} [\href{https://arxiv.org/abs/2311.00753}{{\ttfamily
			2311.00753}}].
	
	\bibitem{Craps:2024suj}
	B.~Craps, O.~Evnin and G.~Pascuzzi, \emph{{Multiseed Krylov complexity}},
	\href{https://arxiv.org/abs/2409.15666}{{\ttfamily 2409.15666}}.
	
	\bibitem{Camargo:2022rnt}
	H.A.~Camargo, V.~Jahnke, K.-Y.~Kim and M.~Nishida, \emph{{Krylov complexity in
			free and interacting scalar field theories with bounded power spectrum}},
	\href{https://doi.org/10.1007/JHEP05(2023)226}{\emph{JHEP} {\bfseries 05}
		(2023) 226} [\href{https://arxiv.org/abs/2212.14702}{{\ttfamily
			2212.14702}}].
	
	\bibitem{He:2024xjp}
	P.-Z.~He and H.-Q.~Zhang, \emph{{Probing Krylov complexity in scalar field
			theory with general temperatures}},
	\href{https://doi.org/10.1007/JHEP11(2024)014}{\emph{JHEP} {\bfseries 11}
		(2024) 014} [\href{https://arxiv.org/abs/2407.02756}{{\ttfamily
			2407.02756}}].
	
	\bibitem{Chattopadhyay:2024pdj}
	A.~Chattopadhyay, V.~Malvimat and A.~Mitra, \emph{{Krylov complexity of
			deformed conformal field theories}},
	\href{https://doi.org/10.1007/JHEP08(2024)053}{\emph{JHEP} {\bfseries 08}
		(2024) 053} [\href{https://arxiv.org/abs/2405.03630}{{\ttfamily
			2405.03630}}].
	
	\bibitem{Malvimat:2024vhr}
	V.~Malvimat, S.~Porey and B.~Roy, \emph{{Krylov Complexity in $2d$ CFTs with
			SL$(2,\mathbb{R})$ deformed Hamiltonians}},
	\href{https://arxiv.org/abs/2402.15835}{{\ttfamily 2402.15835}}.
	
	\bibitem{Vasli:2023syq}
	M.J.~Vasli, K.~Babaei~Velni, M.R.~Mohammadi~Mozaffar, A.~Mollabashi and
	M.~Alishahiha, \emph{{Krylov complexity in Lifshitz-type scalar field
			theories}}, \href{https://doi.org/10.1140/epjc/s10052-024-12609-9}{\emph{Eur.
			Phys. J. C} {\bfseries 84} (2024) 235}
	[\href{https://arxiv.org/abs/2307.08307}{{\ttfamily 2307.08307}}].
	
	\bibitem{Kundu:2023hbk}
	A.~Kundu, V.~Malvimat and R.~Sinha, \emph{{State dependence of Krylov
			complexity in 2d CFTs}},
	\href{https://doi.org/10.1007/JHEP09(2023)011}{\emph{JHEP} {\bfseries 09}
		(2023) 011} [\href{https://arxiv.org/abs/2303.03426}{{\ttfamily
			2303.03426}}].
	
	\bibitem{Avdoshkin:2022xuw}
	A.~Avdoshkin, A.~Dymarsky and M.~Smolkin, \emph{{Krylov complexity in quantum
			field theory, and beyond}},
	\href{https://doi.org/10.1007/JHEP06(2024)066}{\emph{JHEP} {\bfseries 06}
		(2024) 066} [\href{https://arxiv.org/abs/2212.14429}{{\ttfamily
			2212.14429}}].
	
	\bibitem{Khetrapal:2022dzy}
	S.~Khetrapal, \emph{{Chaos and operator growth in 2d CFT}},
	\href{https://doi.org/10.1007/JHEP03(2023)176}{\emph{JHEP} {\bfseries 03}
		(2023) 176} [\href{https://arxiv.org/abs/2210.15860}{{\ttfamily
			2210.15860}}].
	
	\bibitem{Adhikari:2022whf}
	K.~Adhikari, S.~Choudhury and A.~Roy, \emph{{Krylov Complexity in Quantum Field
			Theory}}, \href{https://doi.org/10.1016/j.nuclphysb.2023.116263}{\emph{Nucl.
			Phys. B} {\bfseries 993} (2023) 116263}
	[\href{https://arxiv.org/abs/2204.02250}{{\ttfamily 2204.02250}}].
	
	\bibitem{Dymarsky:2021bjq}
	A.~Dymarsky and M.~Smolkin, \emph{{Krylov complexity in conformal field
			theory}}, \href{https://doi.org/10.1103/PhysRevD.104.L081702}{\emph{Phys.
			Rev. D} {\bfseries 104} (2021) L081702}
	[\href{https://arxiv.org/abs/2104.09514}{{\ttfamily 2104.09514}}].
	
	\bibitem{Rabinovici:2023yex}
	E.~Rabinovici, A.~S\'anchez-Garrido, R.~Shir and J.~Sonner, \emph{{A bulk
			manifestation of Krylov complexity}},
	\href{https://doi.org/10.1007/JHEP08(2023)213}{\emph{JHEP} {\bfseries 08}
		(2023) 213} [\href{https://arxiv.org/abs/2305.04355}{{\ttfamily
			2305.04355}}].
	
	\bibitem{Kar:2021nbm}
	A.~Kar, L.~Lamprou, M.~Rozali and J.~Sully, \emph{{Random matrix theory for
			complexity growth and black hole interiors}},
	\href{https://doi.org/10.1007/JHEP01(2022)016}{\emph{JHEP} {\bfseries 01}
		(2022) 016} [\href{https://arxiv.org/abs/2106.02046}{{\ttfamily
			2106.02046}}].
	
	\bibitem{Adhikari:2022oxr}
	K.~Adhikari and S.~Choudhury, \emph{{Cosmological Krylov Complexity}},
	\href{https://doi.org/10.1002/prop.202200126}{\emph{Fortsch. Phys.}
		{\bfseries 70} (2022) 2200126}
	[\href{https://arxiv.org/abs/2203.14330}{{\ttfamily 2203.14330}}].
	
	\bibitem{Li:2024ljz}
	T.~Li and L.-H.~Liu, \emph{{Krylov complexity of thermal state in early
			universe}},  \href{https://arxiv.org/abs/2408.03293}{{\ttfamily 2408.03293}}.
	
	\bibitem{Li:2024iji}
	T.~Li and L.-H.~Liu, \emph{{Inflationary complexity of thermal state}},
	\href{https://arxiv.org/abs/2405.01433}{{\ttfamily 2405.01433}}.
	
	\bibitem{Li:2024kfm}
	T.~Li and L.-H.~Liu, \emph{{Inflationary Krylov complexity}},
	\href{https://doi.org/10.1007/JHEP04(2024)123}{\emph{JHEP} {\bfseries 04}
		(2024) 123} [\href{https://arxiv.org/abs/2401.09307}{{\ttfamily
			2401.09307}}].
	
	\bibitem{Nandy:2024htc}
	P.~Nandy, A.S.~Matsoukas-Roubeas, P.~Mart\'\i{}nez-Azcona, A.~Dymarsky and
	A.~del Campo, \emph{{Quantum Dynamics in Krylov Space: Methods and
			Applications}},  \href{https://arxiv.org/abs/2405.09628}{{\ttfamily
			2405.09628}}.
	
	\bibitem{Hashimoto:2017oit}
	K.~Hashimoto, K.~Murata and R.~Yoshii, \emph{{Out-of-time-order correlators in
			quantum mechanics}},
	\href{https://doi.org/10.1007/JHEP10(2017)138}{\emph{JHEP} {\bfseries 10}
		(2017) 138} [\href{https://arxiv.org/abs/1703.09435}{{\ttfamily
			1703.09435}}].
	
	\bibitem{PhysRevE.50.888}
	M.~Srednicki, \emph{Chaos and quantum thermalization},
	\href{https://doi.org/10.1103/PhysRevE.50.888}{\emph{Phys. Rev. E} {\bfseries
			50} (1994) 888}.
	
	\bibitem{PhysRevE.99.032213}
	A.~Piga, M.~Lewenstein and J.Q.~Quach, \emph{Quantum chaos and entanglement in
		ergodic and nonergodic systems},
	\href{https://doi.org/10.1103/PhysRevE.99.032213}{\emph{Phys. Rev. E}
		{\bfseries 99} (2019) 032213}.
	
	\bibitem{PhysRevD.100.046020}
	A.R.~Brown and L.~Susskind, \emph{Complexity geometry of a single qubit},
	\href{https://doi.org/10.1103/PhysRevD.100.046020}{\emph{Phys. Rev. D}
		{\bfseries 100} (2019) 046020}.
	
	\bibitem{Lv:2023jbv}
	C.~Lv, R.~Zhang and Q.~Zhou, \emph{{Building Krylov complexity from circuit
			complexity}},
	\href{https://doi.org/10.1103/PhysRevResearch.6.L042001}{\emph{Phys. Rev.
			Res.} {\bfseries 6} (2024) L042001}
	[\href{https://arxiv.org/abs/2303.07343}{{\ttfamily 2303.07343}}].
	
	\bibitem{Balasubramanian:2022tpr}
	V.~Balasubramanian, P.~Caputa, J.M.~Magan and Q.~Wu, \emph{{Quantum chaos and
			the complexity of spread of states}},
	\href{https://doi.org/10.1103/PhysRevD.106.046007}{\emph{Phys. Rev. D}
		{\bfseries 106} (2022) 046007}
	[\href{https://arxiv.org/abs/2202.06957}{{\ttfamily 2202.06957}}].
	
	\bibitem{Ganguli:2024uiq}
	M.~Ganguli, \emph{{Spread Complexity in Non-Hermitian Many-Body Localization
			Transition}},  \href{https://arxiv.org/abs/2411.11347}{{\ttfamily
			2411.11347}}.
	
	\bibitem{Fu:2024fdm}
	Y.~Fu, K.-Y.~Kim, K.~Pal and K.~Pal, \emph{{Statistics and Complexity of
			Wavefunction Spreading in Quantum Dynamical Systems}},
	\href{https://arxiv.org/abs/2411.09390}{{\ttfamily 2411.09390}}.
	
	\bibitem{Nandy:2024mml}
	P.~Nandy, T.~Pathak, Z.-Y.~Xian and J.~Erdmenger, \emph{{A Krylov space
			approach to Singular Value Decomposition in non-Hermitian systems}},
	\href{https://arxiv.org/abs/2411.09309}{{\ttfamily 2411.09309}}.
	
	\bibitem{Fan:2024iop}
	Z.-Y.~Fan, \emph{{Momentum-Krylov complexity correspondence}},
	\href{https://arxiv.org/abs/2411.04492}{{\ttfamily 2411.04492}}.
	
	\bibitem{Xu:2024gfm}
	J.~Xu, \emph{{On Chord Dynamics and Complexity Growth in Double-Scaled SYK}},
	\href{https://arxiv.org/abs/2411.04251}{{\ttfamily 2411.04251}}.
	
	\bibitem{Caputa:2024sux}
	P.~Caputa, B.~Chen, R.W.~McDonald, J.~Sim\'on and B.~Strittmatter,
	\emph{{Spread Complexity Rate as Proper Momentum}},
	\href{https://arxiv.org/abs/2410.23334}{{\ttfamily 2410.23334}}.
	
	\bibitem{Baggioli:2024wbz}
	M.~Baggioli, K.-B.~Huh, H.-S.~Jeong, K.-Y.~Kim and J.F.~Pedraza, \emph{{Krylov
			complexity as an order parameter for quantum chaotic-integrable
			transitions}},  \href{https://arxiv.org/abs/2407.17054}{{\ttfamily
			2407.17054}}.
	
	\bibitem{Rabinovici:2020ryf}
	E.~Rabinovici, A.~S\'anchez-Garrido, R.~Shir and J.~Sonner, \emph{{Operator
			complexity: a journey to the edge of Krylov space}},
	\href{https://doi.org/10.1007/JHEP06(2021)062}{\emph{JHEP} {\bfseries 06}
		(2021) 062} [\href{https://arxiv.org/abs/2009.01862}{{\ttfamily
			2009.01862}}].
	
	\bibitem{Bhattacharjee:2022vlt}
	B.~Bhattacharjee, X.~Cao, P.~Nandy and T.~Pathak, \emph{{Krylov complexity in
			saddle-dominated scrambling}},
	\href{https://doi.org/10.1007/JHEP05(2022)174}{\emph{JHEP} {\bfseries 05}
		(2022) 174} [\href{https://arxiv.org/abs/2203.03534}{{\ttfamily
			2203.03534}}].
	
	\bibitem{Rabinovici:2021qqt}
	E.~Rabinovici, A.~S\'anchez-Garrido, R.~Shir and J.~Sonner, \emph{{Krylov
			localization and suppression of complexity}},
	\href{https://doi.org/10.1007/JHEP03(2022)211}{\emph{JHEP} {\bfseries 03}
		(2022) 211} [\href{https://arxiv.org/abs/2112.12128}{{\ttfamily
			2112.12128}}].
	
	\bibitem{Caputa:2022eye}
	P.~Caputa and S.~Liu, \emph{{Quantum complexity and topological phases of
			matter}}, \href{https://doi.org/10.1103/PhysRevB.106.195125}{\emph{Phys. Rev.
			B} {\bfseries 106} (2022) 195125}
	[\href{https://arxiv.org/abs/2205.05688}{{\ttfamily 2205.05688}}].
	
	\bibitem{Bhattacharya:2023zqt}
	A.~Bhattacharya, P.~Nandy, P.P.~Nath and H.~Sahu, \emph{{On Krylov complexity
			in open systems: an approach via bi-Lanczos algorithm}},
	\href{https://doi.org/10.1007/JHEP12(2023)066}{\emph{JHEP} {\bfseries 12}
		(2023) 066} [\href{https://arxiv.org/abs/2303.04175}{{\ttfamily
			2303.04175}}].
	
	\bibitem{Erdmenger:2023wjg}
	J.~Erdmenger, S.-K.~Jian and Z.-Y.~Xian, \emph{{Universal chaotic dynamics from
			Krylov space}}, \href{https://doi.org/10.1007/JHEP08(2023)176}{\emph{JHEP}
		{\bfseries 08} (2023) 176}
	[\href{https://arxiv.org/abs/2303.12151}{{\ttfamily 2303.12151}}].
	
	\bibitem{Bhattacharjee:2022qjw}
	B.~Bhattacharjee, S.~Sur and P.~Nandy, \emph{{Probing quantum scars and weak
			ergodicity breaking through quantum complexity}},
	\href{https://doi.org/10.1103/PhysRevB.106.205150}{\emph{Phys. Rev. B}
		{\bfseries 106} (2022) 205150}
	[\href{https://arxiv.org/abs/2208.05503}{{\ttfamily 2208.05503}}].
	
	\bibitem{Bhattacharya:2022gbz}
	A.~Bhattacharya, P.~Nandy, P.P.~Nath and H.~Sahu, \emph{{Operator growth and
			Krylov construction in dissipative open quantum systems}},
	\href{https://doi.org/10.1007/JHEP12(2022)081}{\emph{JHEP} {\bfseries 12}
		(2022) 081} [\href{https://arxiv.org/abs/2207.05347}{{\ttfamily
			2207.05347}}].
	
	\bibitem{Bhattacharjee:2022lzy}
	B.~Bhattacharjee, X.~Cao, P.~Nandy and T.~Pathak, \emph{{Operator growth in
			open quantum systems: lessons from the dissipative SYK}},
	\href{https://doi.org/10.1007/JHEP03(2023)054}{\emph{JHEP} {\bfseries 03}
		(2023) 054} [\href{https://arxiv.org/abs/2212.06180}{{\ttfamily
			2212.06180}}].
	
	\bibitem{Camargo:2023eev}
	H.A.~Camargo, V.~Jahnke, H.-S.~Jeong, K.-Y.~Kim and M.~Nishida, \emph{{Spectral
			and Krylov complexity in billiard systems}},
	\href{https://doi.org/10.1103/PhysRevD.109.046017}{\emph{Phys. Rev. D}
		{\bfseries 109} (2024) 046017}
	[\href{https://arxiv.org/abs/2306.11632}{{\ttfamily 2306.11632}}].
	
	\bibitem{Huh:2023jxt}
	K.-B.~Huh, H.-S.~Jeong and J.F.~Pedraza, \emph{{Spread complexity in
			saddle-dominated scrambling}},
	\href{https://doi.org/10.1007/JHEP05(2024)137}{\emph{JHEP} {\bfseries 05}
		(2024) 137} [\href{https://arxiv.org/abs/2312.12593}{{\ttfamily
			2312.12593}}].
	
	\bibitem{Camargo:2024deu}
	H.A.~Camargo, K.-B.~Huh, V.~Jahnke, H.-S.~Jeong, K.-Y.~Kim and M.~Nishida,
	\emph{{Spread and spectral complexity in quantum spin chains: from
			integrability to chaos}},
	\href{https://doi.org/10.1007/JHEP08(2024)241}{\emph{JHEP} {\bfseries 08}
		(2024) 241} [\href{https://arxiv.org/abs/2405.11254}{{\ttfamily
			2405.11254}}].
	
	\bibitem{He:2022ryk}
	S.~He, P.H.C.~Lau, Z.-Y.~Xian and L.~Zhao, \emph{{Quantum chaos, scrambling and
			operator growth in $ T\overline{T} $ deformed SYK models}},
	\href{https://doi.org/10.1007/JHEP12(2022)070}{\emph{JHEP} {\bfseries 12}
		(2022) 070} [\href{https://arxiv.org/abs/2209.14936}{{\ttfamily
			2209.14936}}].
	
	\bibitem{Caputa:2024vrn}
	P.~Caputa, H.-S.~Jeong, S.~Liu, J.F.~Pedraza and L.-C.~Qu, \emph{{Krylov
			complexity of density matrix operators}},
	\href{https://doi.org/10.1007/JHEP05(2024)337}{\emph{JHEP} {\bfseries 05}
		(2024) 337} [\href{https://arxiv.org/abs/2402.09522}{{\ttfamily
			2402.09522}}].
	
	\bibitem{Hornedal:2022pkc}
	N.~H\"ornedal, N.~Carabba, A.S.~Matsoukas-Roubeas and A.~del Campo,
	\emph{{Ultimate Speed Limits to the Growth of Operator Complexity}},
	\href{https://doi.org/10.1038/s42005-022-00985-1}{\emph{Commun. Phys.}
		{\bfseries 5} (2022) 207} [\href{https://arxiv.org/abs/2202.05006}{{\ttfamily
			2202.05006}}].
	
	\bibitem{Bento:2023bjn}
	P.H.S.~Bento, A.~del Campo and L.C.~C\'eleri, \emph{{Krylov complexity and
			dynamical phase transition in the quenched Lipkin-Meshkov-Glick model}},
	\href{https://doi.org/10.1103/PhysRevB.109.224304}{\emph{Phys. Rev. B}
		{\bfseries 109} (2024) 224304}
	[\href{https://arxiv.org/abs/2312.05321}{{\ttfamily 2312.05321}}].
	
	\bibitem{Nandy:2023brt}
	S.~Nandy, B.~Mukherjee, A.~Bhattacharyya and A.~Banerjee, \emph{{Quantum state
			complexity meets many-body scars}},
	\href{https://doi.org/10.1088/1361-648X/ad1a7b}{\emph{J. Phys. Condens.
			Matter} {\bfseries 36} (2024) 155601}
	[\href{https://arxiv.org/abs/2305.13322}{{\ttfamily 2305.13322}}].
	
	\bibitem{Dymarsky:2019elm}
	A.~Dymarsky and A.~Gorsky, \emph{{Quantum chaos as delocalization in Krylov
			space}}, \href{https://doi.org/10.1103/PhysRevB.102.085137}{\emph{Phys. Rev.
			B} {\bfseries 102} (2020) 085137}
	[\href{https://arxiv.org/abs/1912.12227}{{\ttfamily 1912.12227}}].
	
	\bibitem{altland2010condensed}
	A.~Altland and B.D.~Simons, \emph{Condensed matter field theory}, Cambridge
	university press (2010).
	
	\bibitem{harris2014pedestrian}
	E.G.~Harris, \emph{A pedestrian approach to quantum field theory}, Courier
	Corporation (2014).
	
	\bibitem{Mintchev:2022xqh}
	M.~Mintchev, D.~Pontello, A.~Sartori and E.~Tonni, \emph{{Entanglement
			entropies of an interval in the free Schr\"odinger field theory at finite
			density}}, \href{https://doi.org/10.1007/JHEP07(2022)120}{\emph{JHEP}
		{\bfseries 07} (2022) 120}
	[\href{https://arxiv.org/abs/2201.04522}{{\ttfamily 2201.04522}}].
	
	\bibitem{sakurai2020modern}
	J.J.~Sakurai and J.~Napolitano, \emph{Modern quantum mechanics}, Cambridge
	University Press (2020).
	
	\bibitem{Caputa:2021sib}
	P.~Caputa, J.M.~Magan and D.~Patramanis, \emph{{Geometry of Krylov
			complexity}},
	\href{https://doi.org/10.1103/PhysRevResearch.4.013041}{\emph{Phys. Rev.
			Res.} {\bfseries 4} (2022) 013041}
	[\href{https://arxiv.org/abs/2109.03824}{{\ttfamily 2109.03824}}].
	
	\bibitem{geroch2013quantum}
	R.~Geroch, \emph{Quantum field theory: 1971 lecture notes}, vol.~2, Minkowski
	Institute Press (2013).
	
	\bibitem{viswanath1994recursion}
	V.~Viswanath and G.~M{\"u}ller, \emph{The recursion method: application to many
		body dynamics}, vol.~23, Springer Science \& Business Media (1994).
	
	\bibitem{Avdoshkin:2019trj}
	A.~Avdoshkin and A.~Dymarsky, \emph{{Euclidean operator growth and quantum
			chaos}}, \href{https://doi.org/10.1103/PhysRevResearch.2.043234}{\emph{Phys.
			Rev. Res.} {\bfseries 2} (2020) 043234}
	[\href{https://arxiv.org/abs/1911.09672}{{\ttfamily 1911.09672}}].
	
	\bibitem{Maldacena:2015waa}
	J.~Maldacena, S.H.~Shenker and D.~Stanford, \emph{{A bound on chaos}},
	\href{https://doi.org/10.1007/JHEP08(2016)106}{\emph{JHEP} {\bfseries 08}
		(2016) 106} [\href{https://arxiv.org/abs/1503.01409}{{\ttfamily
			1503.01409}}].
	
	\bibitem{peskin2018introduction}
	M.E.~Peskin, \emph{An introduction to quantum field theory}, CRC press (2018).
	
	\bibitem{kapusta2007finite}
	J.I.~Kapusta and C.~Gale, \emph{Finite-temperature field theory: Principles and
		applications}, Cambridge university press (2007).
	
	\bibitem{Barbon:2019wsy}
	J.L.F.~Barb\'on, E.~Rabinovici, R.~Shir and R.~Sinha, \emph{{On The Evolution
			Of Operator Complexity Beyond Scrambling}},
	\href{https://doi.org/10.1007/JHEP10(2019)264}{\emph{JHEP} {\bfseries 10}
		(2019) 264} [\href{https://arxiv.org/abs/1907.05393}{{\ttfamily
			1907.05393}}].
	
	\bibitem{Muck:2022xfc}
	W.~M\"uck and Y.~Yang, \emph{{Krylov complexity and orthogonal polynomials}},
	\href{https://doi.org/10.1016/j.nuclphysb.2022.115948}{\emph{Nucl. Phys. B}
		{\bfseries 984} (2022) 115948}
	[\href{https://arxiv.org/abs/2205.12815}{{\ttfamily 2205.12815}}].
	
\end{thebibliography}

\providecommand{\href}[2]{#2}\begingroup\raggedright\endgroup

\end{document}